\documentclass[a4paper,11pt]{article}
\usepackage{jheppub} 
\usepackage{lineno}
\usepackage{comment}
\usepackage[most]{tcolorbox}

\title{\boldmath The Drell-Yan Process at NNLL$^\prime$+NNLO using Rapidity Dependent Jet Vetoes}

\author[a]{Thomas Clark,}
\author[b]{Shireen Gangal,}
\author[a]{and Jonathan R. Gaunt.}

\affiliation[a]{Department of Physics and Astronomy, University of Manchester, Manchester M13 9PL, U.K.}
\affiliation[b]{Department of Physics, Ramnarain Ruia Autonomous College, Matunga (E), Mumbai, India.}
\emailAdd{thomas.clark-3@manchester.ac.uk}
\emailAdd{shireengangal@ruiacollege.edu}
\emailAdd{jonathan.gaunt@manchester.ac.uk}

\abstract{We present results for the Drell-Yan process $pp \to Z/\gamma^* + X \to l^+l^- + X$ in the presence of the rapidity dependent jet vetoes $\Tau_{Bj}$ and $\Tau_{Cj}.$ These observables provide a tighter veto at central rapidities than at forward rapidities. Our predictions contain a resummation of large logarithms of the veto scale over the hard scale, matched to fixed order in perturbation theory; we present results at both NLL$^\prime$+NLO and NNLL$^\prime$+NNLO. Uncertainty estimates are provided that contain both resummation and fixed-order uncertainties; we find that using a standard profile scale variation procedure for the former seems to underestimate the uncertainty, particularly at NLL$^\prime$+NLO. Consequently, we modify the procedure to avoid cancellation of scale variations between partonic channels, and verify that the uncertainties from this method are reasonable using a simplified version of the Theory Nuisance Parameter (TNP) method. We then find that the uncertainty decreases going from the NLL$^\prime$+NLO to the NNLL$^\prime$+NNLO predictions, and obtain an uncertainty at the level of $1-3\%$ in the  NNLL$^\prime$+NNLO predictions when the veto scale is of $\mathcal{O}(10 \text{ GeV})$.
}

\newcommand{\Tau}{\mathcal{T}}

\newcommand{\Ecm}{E_\mathrm{cm}}

\newcommand{\cut}{\mathrm{cut}}

\newcommand{\nons}{\mathrm{nons}}
\newcommand{\resum}{\mathrm{resum}}

\begin{document}
\maketitle
\flushbottom

\section{\label{sec:intro}Introduction}

Jet vetoes are an important class of jet-based observable that can be used to cut away background events and separate different hard scattering processes that produce differing numbers of hadronic jets. Often, `tight' cuts, where the jet veto scale $\mathcal{T^{\mathrm{cut}}}$ is much smaller than the hard scale $Q$, are required to efficiently remove background processes. In these circumstances large logarithms of $Q/\mathcal{T^{\mathrm{cut}}}$ appear in the perturbative series for the observable, that need to be summed to all orders to obtain a reliable prediction \cite{Stewart:2009yx,Berger:2010xi}. 

The most commonly used jet veto variable is the transverse momentum of the leading jet (see refs.\@ \cite{Banfi:2012yh, Banfi:2012jm, Becher:2012qa, Becher:2013xia, Stewart:2013faa, Banfi:2015pju, Abreu:2022zgo, Abreu:2022sdc, Bell:2022nrj, Bell:2024epn, Campbell:2023cha} for theoretical calculations for this observable in the context of various processes). Although this is a very useful observable, it is uniform in (pseudo)rapidity space. In experiments, it can be difficult to disentangle low $p_{Tj}$ pile up jets from those from the primary proton-proton collision at high rapidity, due to missing tracking information. Thus, one may want, or need, to use an observable where the cut on $p_{Tj}$ is looser at higher rapidities. One option is to simply switch to a looser cut beyond some particular pseudorapidity value; the 
structure of theoretical predictions for such step-like vetoes is discussed in ref.\@ \cite{Michel:2018hui}. An alternative is to smoothly relax the veto as one goes further forward \cite{Tackmann:2012bt, Gangal:2014qda}. Two examples of such an observable are:
\begin{align}
   \mathcal{T}_{Bj} &= m_{Tj} e^{-\mid y_{j}-Y \mid},
   \label{eq:TauB}
   \\
   \mathcal{T}_{Cj} &= \dfrac{m_{Tj}}{2\cosh\left({y_{j}-Y}\right)},
   \label{eq:TauC}
\end{align}
where $m_{Tj}^2 = m_j^2 + p_{Tj}^2$ and $Y$ is the rapidity of the hard system under study (for Drell-Yan this will be the rapidity of the lepton pair). The structure of the resummation for colour singlet 0-jet processes with a $\mathcal{T}_{B/Cj}$ veto was established in ref.\@ \cite{Tackmann:2012bt}. For the gluon-fusion Higgs cross section, predictions for the 0-jet cross section with a $\mathcal{T}_{B/Cj}$ veto have been obtained at NLL$^\prime$ + NLO \cite{Gangal:2014qda}, and NNLL$^\prime$ + NNLO \cite{Gangal:2020qik}. Experimentally, the cross-section differential in $\mathcal{T}_{Cj}$ has been measured in the $H \to \gamma \gamma$ process \cite{ATLAS:2014yga, ATLAS:2022fnp}, and cross-sections differential in both $\mathcal{T}_{Bj}$ and $\mathcal{T}_{Cj}$ were measured in the context of the $H \to ZZ$ process \cite{CMS:2023gjz}.

In this paper we focus on obtaining predictions for the Drell-Yan process, $pp \rightarrow Z/\gamma^* + X \rightarrow l^+ l^- + X$, in the presence of a $\mathcal{T}_{B/Cj}$ veto. The Drell-Yan process is a valuable standard candle process in QCD, having high rate and clean experimental signature. For this process, fixed-order predictions have been obtained up to N$^3$LO for the total cross section \cite{Duhr:2020seh, Duhr:2021vwj} and the rapidity distribution \cite{Chen:2021vtu}, and (for example) resummed predictions have been obtained for the transverse momentum of the $l^+ l^-$ up to N$^3$LL/N$^4$LL \cite{Bizon:2018foh, Bacchetta:2019sam, Ebert:2020dfc, Alioli:2021qbf, Camarda:2021ict, Ju:2021lah, Neumann:2022lft, Moos:2023yfa, Camarda:2023dqn,Billis:2024dqq}) and $p_{Tj}$ up to NNLL$^\prime$ \cite{Banfi:2012yh, Becher:2014aya, Campbell:2023cha}, with these predictions being extensively compared to the experimental data. A detailed comparison of the predictions for $\sigma(\mathcal{T}_{B/Cj} < \mathcal{T}^\mathrm{cut})$ obtained here to data will constitute a further valuable test of our understanding of initial-state QCD radiation, since $\mathcal{T}_{B/Cj}$ vetoes partition the radiation phase space in a rather different way to $p_{Tj}$ (technically, they are SCET$_{\mathrm{I}}$ observables whilst $p_{Tj}$ is SCET$_{\mathrm{II}}$). 

A further motivation for studying the Drell-Yan process in the presence of a jet veto is that this kind of process can be a useful probe of the underlying event (UE) in proton-proton collisions. One expects the jet-vetoed cross section to be significantly affected at small $\Tau^\cut$ by the UE; in section 5 of ref.~\cite{Gangal:2020qik} we illustrated (in the context of Higgs production, using $\text{M{\scriptsize AD}G{\scriptsize RAPH}5\_{\scriptsize A}MC@NLO}$+Pythia8 simulations) that this  is indeed the case when $\Tau^\cut $ is a few GeV for all three jet vetoes $\Tau_{Bj}, \Tau_{Cj}$ and $p_{Tj}$. In that study we found that the different jet vetoes were affected to different extents by UE and hadronisation effects; whilst the impact of both effects was smaller for $\Tau_{Bj/Cj}$ than $p_{Tj}$, the impact of the UE relative to hadronisation seemed to be larger for $\Tau_{Bj/Cj}$ than $p_{Tj}$, suggesting the $\Tau_{Bj/Cj}$ might be a `purer' probe of UE effects. A further advantage of using $\Tau_{Bj/Cj}$ to study the UE is the already-mentioned fact that this observable is less affected by pile-up jets at high rapidity (this should be a relevant issue for UE studies where a fairly tight veto is required). The Drell-Yan process is the obvious context in which to make such studies, as the high statistics allows precise experimental measurements, and on the theory side we can obtain precise baseline predictions without UE (see below), enabling a precise extraction of the UE effects. The idea of using the Drell-Yan process with a jet veto/cut to study UE has been suggested previously in several papers, see e.g.~Refs~\cite{Procura:2014cba, Alioli:2016wqt, Bansal:2016iri, PaktinatMehdiabadi:2019ujl,Andersen:2023hzm} (in some of these studies a cut on the $p_T$ of the $l^+l^-$ system is also suggested to enhance the sensitivity to the UE). Here we will not try to model the UE effects, focussing on obtaining the baseline prediction without UE mentioned previously.

In this paper we obtain predictions at both NLL$^\prime$ + NLO and NNLL$^\prime$ + NNLO. We include the resummation of time-like logarithms in the hard process to all orders \cite{Ahrens:2009cxz,Ahrens:2008qu,Ebert:2017uel}, but do not perform resummation of logarithms of the jet radius $R$ (studied in refs.\@ \cite{Dasgupta:2014yra,Kang:2016mcy,Dai:2016hzf}).

The structure of this paper is as follows. In section \ref{sec:resummation}, the factorisation of the Drell-Yan cross section with a $\Tau_{B/Cj}$ veto \cite{Tackmann:2012bt,Gangal:2014qda} is reviewed, as well as the matching of the resummed predictions to fixed order. In section \ref{sec:uncertainty} we describe how we set the various scales in our calculation, and review a procedure to estimate the resummation and fixed order uncertainties in our predictions, following (for example) ref.~\cite{Gangal:2020qik}. We find that using this `standard' procedure, the uncertainty bands for the  NLL$^\prime$ + NLO predictions in particular are rather small, smaller even than the NNLL$^\prime$ + NNLO bands at very small $\Tau^\cut$. We examine why this is in section \ref{sec:convergence}, and propose an alternative `\textit{MaxDev}' approach for estimating the resummation uncertainties that is more conservative and gives more reliable results. We validate these uncertainty estimates using a simplified version of the Theory Nuisance Parameter method proposed in ref.\@ \cite{Tackmann:2024kci}. Our results for the Drell-Yan cross section with a $\Tau_{B/Cj}$ veto are given at NLL$^\prime$ + NLO and NNLL$^\prime$ + NNLO in section \ref{sec:results}, using the \textit{MaxDev} approach to estimate the uncertainties. We conclude in section \ref{sec:conclusion}.
\section{\label{sec:resummation}Factorisation and Resummation of the Drell-Yan Cross Section}

The full Drell-Yan $pp \to Z/\gamma^* + X\to l^+ l^- + X$ cross section with a veto on the rapidity-dependent observable $\Tau_{fj} < \Tau^\cut$ ($f=B,C$) is given by,
\begin{align} \label{eq:fullXsec}
\frac{d\sigma_0 (\Tau_{fj} \!< \Tau^\cut)}{dQ^2 dY}
&= \frac{d\sigma_0^\resum (\Tau_{fj} \!< \Tau^\cut)}{dQ^2 dY}
+ \frac{d\sigma_0^\nons (\Tau_{fj} \!< \Tau^\cut)}{dQ^2 dY}
\,,\end{align}
\noindent where the first term contains the resummed logarithms of $\Tau^\cut/Q$, and dominates at small values of $\Tau^\cut$, while the second term contains the nonsingular corrections which are suppressed by $O(\Tau^\cut/Q)$ and are required to ensure the sum at large $\Tau^\cut$ reproduces the fixed-order result. The resummed cross section can be calculated from a SCET factorisation formula as defined in refs.\@ \cite{Stewart:2009yx, Gangal:2014qda}, 
\begin{align}
\label{eq:TauBCfacto}
\dfrac{d\sigma_0^\resum (\Tau_{fj} \!< \Tau^\cut)}{ dQ^2 dY} =  \, &\sigma_B \sum_{ij} H_{ij}(Q^2, \mu_H)\, U_H(Q^2, \mu_H, \mu)
\\ \nonumber &\times
B_i(Q \Tau^{\cut},x_a,R,\mu_B)
B_j(Q \Tau^{\cut},x_b,R,\mu_B) U_B^2(Q \Tau^{\cut}, R, \mu_B, \mu)
\\ \nonumber &\times S_{f}(\Tau^{\cut},R,\mu_S) U_S( \Tau^{\cut}, R, \mu_S, \mu)
\\ \nonumber & +\frac{d \sigma_0^{\text{Rsub}}(\Tau_{fj} \!< \Tau^\cut, R)}{dQ^2 dY}
\,,
\end{align}
%
\noindent where
\begin{align}
x_{a,b} = \frac{Q}{\Ecm}\,e^{\pm Y}
\,,\quad
\sigma_B = \frac{4\pi \alpha_{em}^2}{3 N_c \Ecm^2 Q^2}
\,,\quad
Q = \sqrt{m_{l^+l^-}^2}
\,.\end{align}
and the sum is over $ij = \{ u\bar{u}, d\bar{d}, c\bar{c}, s\bar{s}, b\bar{b}, \bar{u}u, ... \}$. 
\noindent The factorized formula includes a hard function $H$, beam functions $B$ and a soft function $S$. To resum large (double) logarithms of $\mathcal{T}^\mathrm{cut}/Q$, the hard, beam and soft functions are first evaluated at their `natural' scales $|\mu_H| \sim Q$, $\mu_B \sim \sqrt{Q \Tau^\cut}$ and $\mu_S \sim \Tau^\cut$ where large logarithms are absent from the perturbative series for each individual ingredient. They are then renormalisation group (RG) evolved to a common scale $\mu$ via the respective evolution factors $U_H, U_B$ and $U_S$ which sums up the large logarithms. Here we take this common scale to be the `fixed order' scale  $\mu_{\mathrm{FO}}$, which should be chosen to be of order of the hard process scale $\mu_{\mathrm{FO}} \sim Q$. Ultimately we will produce predictions only for a small range of $Q$ values centred on $M_Z$, so for the purpose of scale setting we will replace $Q$ by $M_Z$ and take the central value of $\mu_{\mathrm{FO}}$ to be $M_Z$.

For NLL$^\prime$ predictions, one needs the fixed-order expansion of $B,H,S$ and (non-cusp) anomalous dimensions of these quantities to one loop, whilst for NNLL$^\prime$ we need all of these quantities at two loops. For the Drell-Yan process, we obtained the one- and two-loop coefficients for the hard function from refs.~\cite{Stewart:2009yx} and \cite{Idilbi:2006dg} respectively. We require the quark beam function $B_q$, which at perturbative $\Tau^\cut$ can be written as the convolution of perturbative matching coefficients $\mathcal{I}_{qj}$ and the standard parton distribution functions (PDFs) \cite{Stewart:2009yx}. The one-loop and two-loop matching coefficients can be obtained by taking the cumulant of the virtuality-dependent quark beam function matching coefficients computed in ref.~\cite{Stewart:2010qs, Gaunt:2014xga}, with an additional contribution at two loops from jet-radius dependent corrections as computed in ref.\@ \cite{Gangal:2016kuo}. Up to two loops, the soft function is similar to the one used for Higgs production given in equations (2.10)-(2.16) of ref.~\cite{Gangal:2020qik} (which uses results from refs.~\cite{Monni:2011gb, Kelley:2011ng, Hornig:2011iu, Hoang:2014wka, Gaunt:2015pea, Gangal:2016kuo}, see also ref.~\cite{Bell:2018oqa}). The only differences are that in $S_f^{(1)}$ one replaces $C_A \to C_F$, in the expression for $S_{G,f}^{(2),\text{ non-Ab}}$ in equation (2.12) one replaces the overall prefactor of $C_A$ by $C_F$, and for $\Delta S_f^{(2)}$ one uses the correction appropriate for the quark case from section 3.1.1 of ref.~\cite{Gangal:2016kuo} rather than that for the gluon case. The two-loop noncusp anomalous dimensions for the beam, soft and hard functions can be constructed from equations (3.29) and (3.5) of ref.~\cite{Gangal:2020qik} and Appendix D of ref.~\cite{Stewart:2010qs}. 

\begin{figure}[!htb]
\centering
\includegraphics[width=.75\textwidth]{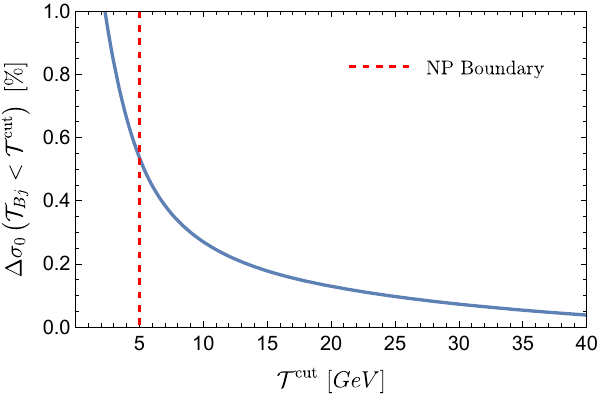}
\caption{Percent difference in the NNLL$^\prime$ $\mathcal{T}_{Bj} \leq \mathcal{T}^{\text{cut}}$ on-shell $Z$ boson production cross section when using different treatments for the $O(R^2)$ corrections associated with the clustering of independent emissions.}
\label{fig:indepClustering}
\end{figure}

The $d\sigma_0^{Rsub}$ term in eq.~\eqref{eq:TauBCfacto} contains $O(R^2)$ corrections arising due to clustering of independent emissions.
Discussion on how to treat these terms can be found in refs.\@ \cite{Tackmann:2012bt, Gangal:2016kuo} and leads to two different prescriptions. In this analysis, we followed refs.~\cite{Stewart:2013faa, Gangal:2020qik} and included these independent emission pieces separately from the rest of the resummed factorisation formula. We have $d\sigma_0^{Rsub}=0$ for the NLL$^\prime$ case, and in the NNLL$^\prime$ case the result is given in equation (2.22) of ref.\@ \cite{Gangal:2016kuo} with $g \to q$. At NNLL$^\prime$ we found the difference between the two different prescriptions to be under $1\%$ at $R = 0.5$ outside of the non-perturbative region of our calculation (as defined in section~\ref{sec:uncertainty}) and of $\mathcal{O}\left(0.1\%\right)$ for the majority of the values of $\mathcal{T^{\rm{cut}}}$ as can be seen in figure \ref{fig:indepClustering}.

Let us now discuss the non-singular piece in \eqref{eq:fullXsec}. Here we will only present the explicit formulae for the NNLL$^\prime$ + NNLO case; for the NLL$^\prime$+NLO case one just removes one `N' from all the formulae below. The expression is: 
\begin{multline}
\sigma^{\text{nons,NNLO}}_{0}\left(\Tau_{fj} <\Tau^{\text{cut}}, \mu_{\rm{FO}}\right) \\ =  \sigma^{\text{FO,NNLO}}_{0}\left(\Tau_{fj} < \Tau^{\text{cut}}\right) -  \sigma^{\text{resum,}\text{NNLL}^{\prime}}_{0}\left(\Tau_{fj} < \Tau^{\text{cut}}, \mu_B = \mu_S = \mu_H = \mu_{\rm{FO}}\right)\,, 
\label{eq:nonsing}
\end{multline}
\noindent where
\begin{equation}
\sigma_0^{\text{FO,NNLO}}\left(\Tau_{fj} <\Tau^{\text{cut}}\right) = \sigma^{\text{FO,NNLO}}_{\geq 0} - \sigma^{\text{FO,NLO}}_{\geq 1}\left(\Tau_{fj} >\Tau^{\text{cut}}\right).
\label{eq:FO below}
\end{equation}

\noindent The second term on the right hand side of eq.~\eqref{eq:nonsing} can be easily obtained using the factorisation formula. In this term the soft, beam and hard scale are all evaluated at a common FO scale and there is no resummation of large logarithms. In eq.~\eqref{eq:FO below}, the first piece is the fixed-order NNLO cross section. We compute this using DYTurbo \cite{Camarda:2019zyx} and perform a cross check by using global slicing \cite{Catani:2007vq, Gaunt:2015pea, Boughezal:2015dva}, with $\Tau_{Bj}$ as the resolution variable; these two predictions deviated by $\mathcal{O}\left(0.1\%\right)$ when using cut values in $\Tau_{Bj}$ of $0.25\text{ GeV}$. This was within the statistical errors of the predictions and consistent with the expected size of the non-singular corrections, which go to $0$ as the cut value in the resolution variable goes to $0$ as seen in figure \ref{fig:IvFNNLO}. The second piece in eq.~\eqref{eq:FO below} is the NLO Drell-Yan + 1 jet cross section with $\Tau_{fj} > \Tau^\cut$; we compute this directly using $\text{M{\scriptsize AD}G{\scriptsize RAPH}5\_{\scriptsize A}MC@NLO}$\cite{Alwall:2014hca, Hirschi:2015iia}.

It has been previously mentioned that we include the resummation of time-like logarithms to all orders in this analysis. This is done by choosing the hard scale $\mu_H = -i\mu_{\rm{FO}}$ \cite{Ahrens:2008qu,Ahrens:2009cxz}. When including this, the non-singular piece needs to be changed as follows \cite{Ebert:2017uel,Stewart:2013faa,Gangal:2020qik},
\begin{multline}
\frac{d\sigma^{\text{nons},\text{NNLO}+\pi^2}_{0}\left(\Tau_{fj} < \Tau^{\text{cut}}\right)}{dQ^2 dY} \\ = \big[\frac{d\sigma^{\text{nons},\text{NNLO}}_{0}\left(\Tau_{fj} < \Tau^{\text{cut}}\right)}{dQ^2 dY} - \frac{\alpha_{s}\left(\mu_{\rm{FO}}\right)C_F\pi^2}{2\pi}\times\frac{d\sigma^{\text{nons},\text{NLO}}_{0}\left(\Tau_{fj} < \Tau^{\text{cut}}\right)}{dQ^2 dY}\big] \\ \times U_H\left(Q^2,-i\mu_{\rm{FO}},\mu_{\rm{FO}}\right).
\label{eq:nonsingpi}
\end{multline}

Finally, we can use eq.~\eqref{eq:nonsingpi} along with the initial definition of the non-singular piece in relation to the FO prediction as defined in eq.~\eqref{eq:nonsing} to calculate the FO + $\pi^2$ prediction as follows,
\begin{multline}
\sigma_0^{\text{FO,NNLO}+\pi^2}\left(\Tau_{fj} <\Tau^{\text{cut}}\right) = \sigma^{\text{nons},\text{NNLO}+\pi^2}_{0}\left(\Tau_{fj} < \Tau^{\text{cut}}\right) + \\ \sigma^{\text{resum,}\text{NNLL}^{\prime}}_{0}\left(\Tau_{fj} < \Tau^{\text{cut}}, \mu_B = \mu_S = \mu_{\rm{FO}}, \mu_H = -i\mu_{\rm{FO}}\right).
\label{eq:FOBelowPi}
\end{multline}
\noindent In section \ref{sec:mainResults} predictions described by Eq.~\eqref{eq:fullXsec} and Eq.~\eqref{eq:FOBelowPi} will be compared.

\section{\label{sec:uncertainty}Profile Scales and Uncertainties}

As already mentioned, at small $\Tau^\cut \ll Q$ (in the `resummation region') the scales $\mu_B, \mu_S$ and $\mu_H$ in \eqref{eq:TauBCfacto} should be set to their canonical values $|\mu_H| \sim Q$, $\mu_B \sim \sqrt{Q \Tau^\cut}$ and $\mu_S \sim \Tau^\cut$ in order to perform the resummation of large logarithms of $\Tau^\cut/Q$. On the other hand, at large $\Tau^\cut \gtrsim Q$ (`fixed-order region') the resummation should be `turned off', and $|\mu_H|, \mu_B$ and $\mu_S$ set to $\mu_{\text{FO}}$ such that the fixed order result is obtained via the combination of the singular and nonsingular cross sections. There should be a smooth transition for $\mu_B, \mu_S$ and $\mu_H$ between these two regimes.

We achieve this using profile scales \cite{Ligeti:2008ac,Abbate:2010xh}, using the same general form of the profile scales as was used in ref.~\cite{Gangal:2020qik}. Concretely, we have $\mu_H = -i \mu_{\mathrm{FO}}$, $\mu_S = \mu_{\mathrm{FO}} f_{run}(\Tau^\cut/M_Z)$ and $\mu_B = \mu_{\mathrm{FO}} \sqrt{f_{run}(\Tau^\cut/M_Z)}$, where
\begin{equation}
f_{\text{run}}\left(x\right) = \begin{cases}
x_0\left[1+\left(2r_s - 1\right)\left(x/x_0\right)^2/4\right] & \text{$x \leq 2x_0$}, \\ r_{s}x & \text{$2x_0 \leq x \leq x_1$}, \\ r_{s} x + \frac{\left(2-r_{s}x_2 - r_{s}x_3\right)\left(x-x_1\right)^2}{2\left(x_2 - x_1\right)\left(x_3 - x_1\right)} & \text{$x_1 \leq x \leq x_2$}, \\ 1 - \frac{\left(2-r_{s}x_1 - r_{s}x_2\right)\left(x-x_{3}\right)^2}{2\left(x_3-x_1\right)\left(x_3 - x_2\right)} & \text{$x_2 \leq x \leq x_3$}, \\ 1 & \text{$x_3 \leq x$}.
\end{cases}
\label{eq:frun}
\end{equation}
In \eqref{eq:frun} the $x_i$ determine the boundaries of the different regions; $\Tau^\cut = 2x_0M_Z$ represents the start of the resummation region, $\Tau^\cut = x_1 M_Z$ the beginning of the `transition' region between the resummation and fixed order regions (where the transition occurs in two segments, $x_1 M_Z \le \Tau^\cut \le x_2 M_Z$ and $x_2 M_Z \le \Tau^\cut \le x_3 M_Z$), and $\Tau^\cut = x_3 M_Z$  the beginning of the fixed order region. Below $2x_0 M_Z$ we have a `non-perturbative region' where we ultimately freeze $\mu_B$ and $\mu_S$ to fixed values $> \Lambda_{QCD}$ as $\Tau^\cut \to 0$ in order to avoid $\alpha_s$ and the PDFs being evaluated at too low scale values. The quantity $r_s$ is a parameter that should be chosen of $\mathcal{O}(1)$  -- following the arguments in ref.~\cite{Gangal:2020qik} we take $r_s=1$ for $\Tau_{Bj}$ and $r_s=2$ for $\Tau_{Cj}$.

\begin{figure}
\centering
\hspace{-2mm}
\includegraphics[width=.5\textwidth]{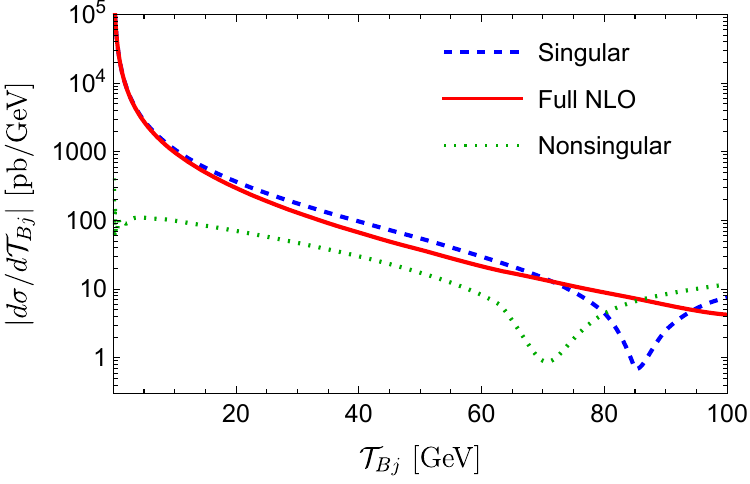}
\hspace{-2mm}
\includegraphics[width=.5\textwidth]{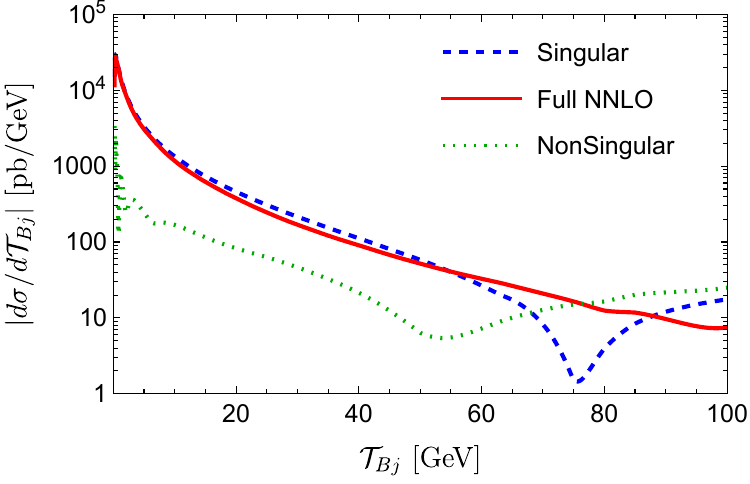}
\caption{The NLO (left) and NNLO (right) on-shell $Z$  cross-section differential in  $\Tau_{Bj}$, split into its singular and non-singular components.}
\label{fig:TauBScaleProfile}
\end{figure}

\begin{figure}
\centering
\hspace{-2mm}
\includegraphics[width=.5\textwidth]{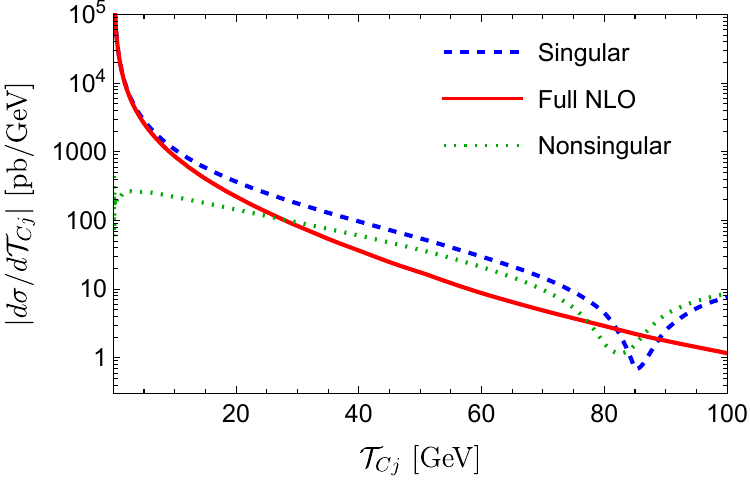}
\hspace{-2mm}
\includegraphics[width=.5\textwidth]{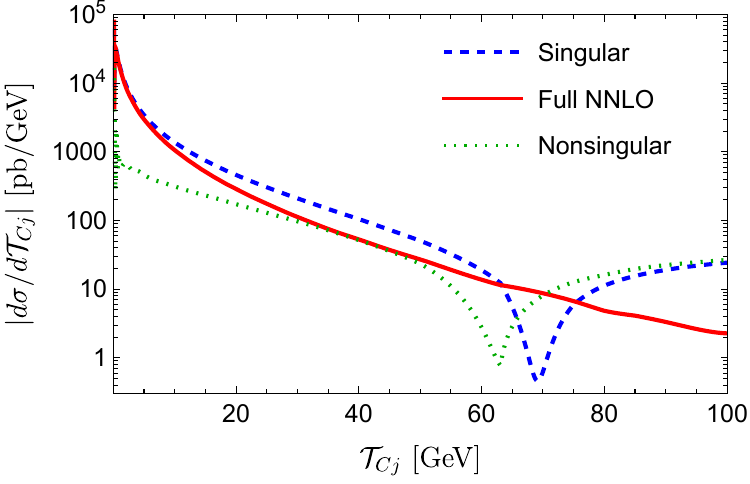}
\caption{The NLO (left) and NNLO (right) on-shell $Z$  cross-section differential in $\Tau_{Cj}$, split into its singular and non-singular components.}
\label{fig:TauCScaleProfile}
\end{figure}

The values of the $x_i$ are determined by comparing the sizes of singular and non-singular contributions at fixed order, as discussed in ref.~\cite{Gangal:2014qda}. In figures~\ref{fig:TauBScaleProfile} and \ref{fig:TauCScaleProfile} we plot the singular, non-singular and full fixed-order cross sections for $\Tau_{Bj}$ and $\Tau_{Cj}$ respectively, for the simplified case of on-shell $Z$ boson production (given that we only consider $Q \sim M_Z$ this is sufficient for the purpose of setting the profile scales). The left hand pane presents the NLO results that can be used to determine the  $x_i$ for the NLL$^\prime + $NLO prediction, whilst the right pane gives the  NNLO results for the NNLL$^\prime + $NNLO prediction. We see that, as expected, the singular cross section is dominant at small $\Tau^\cut$, since the singular cross section diverges as $1/\Tau^\cut$, whilst the non-singular only diverges logarithmically at small $\Tau^\cut$.

We use $x_0 = 2.5/\mu_{\rm{FO}}$ as in ref.~\cite{Stewart:2013faa} to define the non-pertubative region. From the left hand panes of figures \ref{fig:TauBScaleProfile} and \ref{fig:TauCScaleProfile}, we determine that at NLL$^\prime$+NLO, the $x_i$ parameters have the same value \{$x_1$,$x_2$,$x_3$\} = \{$0.20$,$0.55$,$0.90$\} for both $\Tau_{Bj}$ and $\Tau_{Cj}$. At NNLL$^\prime$+NNLO, we find from the right hand panes of figures \ref{fig:TauBScaleProfile} and \ref{fig:TauCScaleProfile} that the values of $x_i$ for $\Tau_{Bj}$ and $\Tau_{Cj}$ should be \{$x_1$,$x_2$,$x_3$\} = \{$0.20$,$0.50$,$0.80$\} and \{$x_1$,$x_2$,$x_3$\} = \{$0.15$,$0.45$,$0.75$\} respectively. Note that our choice of the $x_i$ for $\Tau_{Bj}$ at NNLL$^\prime$+NNLO coincides with the choice made in ref.~\cite{Alioli:2015toa} for beam thrust $\Tau_0$ in Drell-Yan at NNLL$^\prime$+NNLO.

The perturbative uncertainties in our resummed predictions arise from two sources: the overall fixed-order scale variation uncertainty $\Delta_{\rm{FO}}$ and the resummation uncertainty $\Delta_{\rm{resum}}$ which corresponds to the uncertainty in the logarithmic series induced by the jet veto cut.
The fixed-order uncertainty is determined by varying $\mu_{\rm{FO}}$ as \{$\frac{1}{2}M_Z$,$M_Z$,$2M_Z$\}. In order to obtain the resummation uncertainties, the profile scales need to be varied. We use the same function as in ref.~\cite{Gangal:2020qik} to vary the profile scales defined as follows:

\begin{equation}
f_{\text{vary}}\left(x\right) = \begin{cases}
2\left(1-\left(1+\delta\right)x^2/x_3^2\right) & \text{$0 \leq x \leq x_3/2$,} \\ 1 + 2\left(1-3\delta\right)\left(1-x/x_3\right)^2 + 16\delta\left(1-x/x_3\right)^4 & \text{$x_3/2 \leq x \leq x_3$,} \\
1 & \text{$x_3 \leq x$,}
\end{cases}
\label{eq:fvary}
\end{equation}

\noindent where we use $\delta = 0$ for $r_s = 1$, and $\delta = 0.05$ for $r_s = 2$ to ensure that $\mu_B$ and $\mu_S$ scales don't rise above $\mu_{\rm{FO}}$ \cite{Gangal:2020qik}. The beam and soft scales are varied using this multiplicative factor as follows:
\begin{align}
    \mu_{S}^{\text{vary}}\left(x,\alpha\right) &= \mu_{\rm{FO}}f_{\text{vary}}^{\alpha}\left(x\right)f_{\text{run}}\left(x\right), \label{eq:muSVar}\\
    \mu_{B}^{\text{vary}}\left(x,\alpha,\beta\right) &= \mu_{S}^{\text{vary}}\left(x,\alpha\right)^{1/2 - \beta}\mu_{\rm{FO}}^{1/2+\beta} = \mu_{\rm{FO}}\left[f_{\text{vary}}^{\alpha}\left(x\right)f_{\text{run}}\left(x\right)\right]^{1/2 - \beta}.
    \label{eq:muBVar}
\end{align}
\noindent The parameters $\alpha$ and $\beta$ define the variations in the profile scales; for ($\alpha$,$\beta$) = ($0$,$0$) we recover the central profile scale. To obtain $\Delta_{\text{resum}}$ we produce predictions with $(\alpha,\beta) = \{(+1,0),(-1,0),(0,+1/6),(0,-1/6)\}$ and define $\Delta_{\text{resum}}$ as the maximum deviation from the central profile. Note that the factorisation scale $\mu_F$ is always held equal to the beam scale $\mu_B$ in these variations.

The total uncertainty is given by,

\begin{equation}
\Delta_{\text{0}}\left(\Tau^{\text{cut}}\right) = \sqrt{\Delta_{\text{FO}}^2\left(\Tau^{\text{cut}}\right) + \Delta_{\text{resum}}^2\left(\Tau^{\text{cut}}\right)}\,.
\label{eq:overallError}
\end{equation}
\section{\label{sec:convergence}A Closer Look at Perturbative Uncertainties}

\begin{figure}
\centering
\hspace{-2mm}
\includegraphics[width=.5\textwidth]{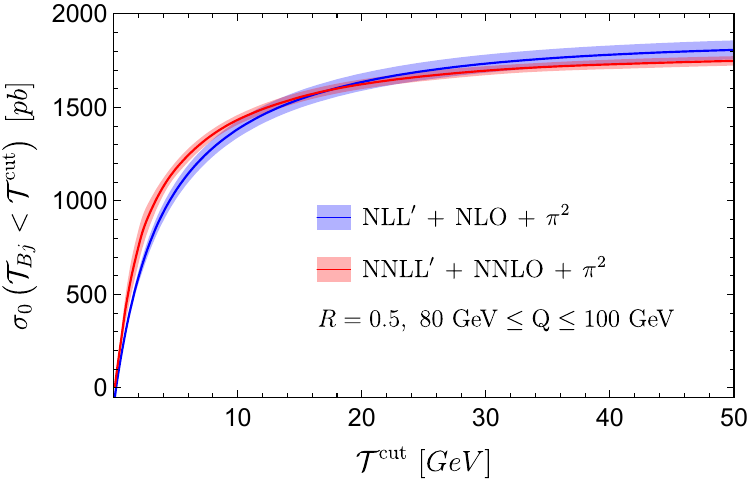}
\hspace{-2mm}
\includegraphics[width=.5\textwidth]{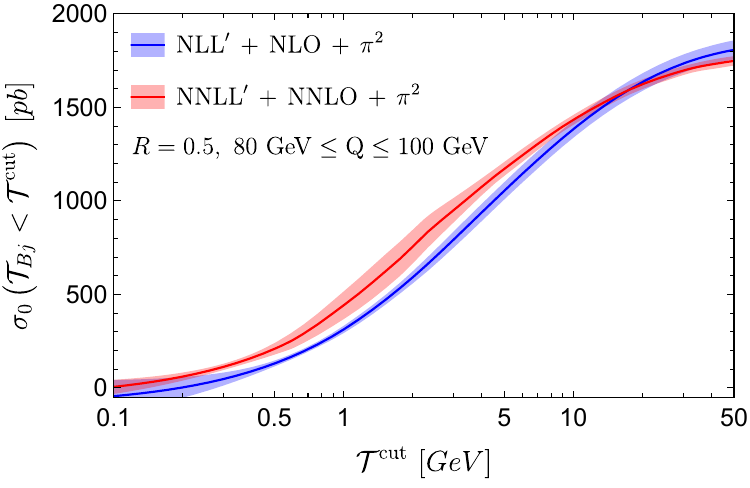}
\caption{Drell-Yan NLL$^\prime$ + NLO (blue) and NNLL$^\prime$ + NNLO (red) cross section for $\Tau_{Bj} < \Tau^\mathrm{cut}$ for $r_s = 1$ plotted on a linear scale (left) and logarithmic scale (right) using standard scale variations.}
\label{fig:tauBstnd}
\end{figure}

Using the procedure for estimating uncertainties described in the previous section, we generated both NLL$^\prime$ + NLO and NNLL$^\prime$ + NNLO results for $\sigma(\Tau_{Bj} < \Tau^\cut)$ with $\sqrt{s} = 13$ TeV, $R = 0.5$, and $Q$ integrated between $80$ and $100$ GeV (the full specification of the set-up is given at the start of section \ref{sec:results}). The result is given in figure \ref{fig:tauBstnd}. In the large $\Tau^\cut$ fixed order region the uncertainty band decreases going from NLL$^\prime$+NLO to NNLL$^\prime$+NNLO and the uncertainty bands overlap, as expected; however the same is not true at small $\Tau^\cut$ values of a few GeV where the bands no longer overlap and the NLL$^\prime$+NLO band eventually becomes smaller than the NNLL$^\prime$+NNLO one. This suggests that at least the resummation uncertainty for the NLL$^\prime$+NLO prediction is being underestimated by the procedure in section \ref{sec:uncertainty}. We have studied in some detail why this is; the remainder of this section summarizes our findings.

A further notable feature of figure \ref{fig:tauBstnd} is the `bulge' in the NNLL$^{\prime}$ + NNLO uncertainty band, particularly noticeable in the plot with a logarithmic scale, when $\Tau^\cut$ is $\mathcal{O}\left(1 \text{ GeV}\right)$. This bulge in the nonperturbative region is linked to the $\alpha$ variations -- in particular the $\alpha = -1$ variation pushing the soft scale to extremely low values, causing large deviations (see figure \ref{fig:softDeviations} below). We shall see in section \ref{sec:softErr} that the response of the NLL$^{\prime}$ prediction to $\mu_s$ variations is unusually small, so such a bulge is not seen in the NLL$^{\prime}$+NLO predictions. In the nonperturbative region we do not expect our predictions to be particularly reliable anyway (we have dropped terms of $\mathcal{O}\left(\Lambda_{\text{QCD}}/\Tau^{\text{cut}}\right)$, that become important in this region), so we will not concern ourselves with this feature any further.

Our study of the scale variation uncertainties is split into two parts, which are given in sections \ref{sec:softErr} and \ref{sec:channelcancellation}; very roughly speaking, these focus on the effect of $\mu_S$ and $\mu_B$ variations respectively. At the end of section \ref{sec:channelcancellation} we give an improved prescription for estimating the theoretical uncertainty, which we refer to as \textit{MaxDev}, that should be more reliable and exhibits better perturbative convergence. In section \ref{sec:Nuisance Parameters}, we use an alternative method of estimating the resummation uncertainties to verify that our \textit{MaxDev} uncertainties are reasonable; this is a simplified version of the theory nuisance parameter method introduced in \cite{Tackmann:2024kci}.

\subsection{\label{sec:softErr}Soft Function Variations}

\begin{figure}[tbp]
\centering
\hspace{-2mm}
\includegraphics[width=.5\textwidth]{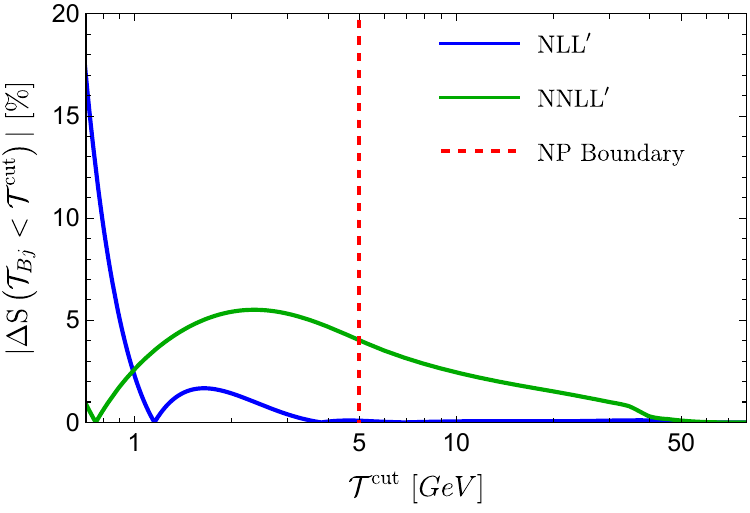}
\hspace{-2mm}
\includegraphics[width=.5\textwidth]{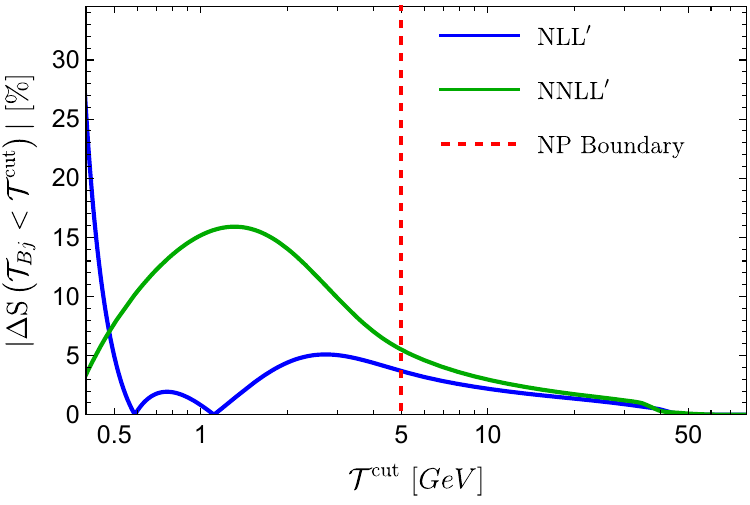}
\caption{Percent deviations in Drell-Yan evolved soft function for $\alpha = 1$ (left) and $\alpha = -1$ (right). The red dashed line marks where we transition into the `non-perturbative region', as defined in section~\ref{sec:uncertainty}.}
\label{fig:softDeviations}
\end{figure}

Here we study the effect of the $\mu_S$ variations in the soft function. Only the $\alpha = \pm 1$ variations need to be discussed as the $\beta$ variations do not vary the soft scale. Let us consider the combination of the soft function $S$ and the soft evolution factor $U_S$ appearing in the resummation formula (the `evolved soft function'):
\begin{equation}
S\left(\Tau^{\text{cut}},R, \mu_S, \mu\right) \equiv S\left(\Tau^{\text{cut}},R,\mu_S\right) U_S\left(\Tau^{\text{cut}}, R,\mu_S,\mu\right),
\label{eq:evoSoft}
\end{equation}
The $\mu_s$ dependence should cancel between $U_S$ and $S$, up to corrections that are of order $\alpha_s^{n+1}$ if $U_S$ and $S$ are evaluated at N$^n$LL$^\prime$. Figure \ref{fig:softDeviations} shows the $\alpha = \pm 1$ deviations in the evolved soft function for $\mu_s$ as described in eq.~\eqref{eq:muSVar} and $\mu = M_Z$. Note the peaking behaviour in the non-perturbative region for the NNLL$^{\prime}$ deviations, which is mostly responsible for the bulge in figure \ref{fig:tauBstnd}.

We see in figure \ref{fig:softDeviations} that the deviation is generally larger for NNLL$^{\prime}$ than NLL$^{\prime}$, which is unexpected and suggests that the coefficient of $\alpha_s^{2}$ for the NLL$^{\prime}$ scale variation is anomalously small (or that of $\alpha_s^{3}$ for the NNLL$^{\prime}$ variation is anomalously large). To investigate this in more detail, let us consider the evolved soft function with $\mu=\Tau^\cut$ -- this then gives us a single scale object, and allows us to focus on the $\mu_s$ dependence dropping the overall Sudakov factor $U_S(\Tau^\cut,\Tau^\cut,\mu)$ between the $\Tau^\cut$ and $\mu$ scales. We expand this in powers of $\alpha_s\left(\mathcal{T}^{\rm{cut}}\right)$ and compute the leading term contributing to the deviation in the evolved $S$ when we vary $\mu_s$ according to eq.~\eqref{eq:muSVar}. We take $\mathcal{T}^{\rm{cut}} = 20 \text{ GeV}$ for this exercise (the pattern of results was similar for other values of $\mathcal{T}^{\rm{cut}}$ in the resummation and early transition regions).

At NLL$^\prime$, the leading term for the $\alpha = 1$ variation is $ 0.07 \alpha^2_s\left(\mathcal{T}^{\rm{cut}}\right)$ whilst that for the $\alpha = -1$ variation is $ 0.43 \alpha^2_s\left(\mathcal{T}^{\rm{cut}}\right)$. However, at NNLL$^\prime$ we find that the leading term for the $\alpha = 1$ variation is $ -4.41 \alpha^3_s\left(\mathcal{T}^{\rm{cut}}\right)$ whilst that for the $\alpha = -1$ variation is $ 3.65 \alpha^3_s\left(\mathcal{T}^{\rm{cut}}\right)$. The coefficients of $\alpha_s^2$ for the NLL$^\prime$ variations are rather small (particularly for $\alpha = 1$), and are much smaller than those of $\alpha_s^3$ for the NNLL$^\prime$ variations. We also checked the subleading terms in $\alpha_s$ for the deviations, and observed that, at a given order in  $\alpha_s\left(\mathcal{T}^{\rm{cut}}\right)$ the size of the coefficients are smaller for the NLL$^{\prime}$ variations than the NNLL$^{\prime}$ ones (checked up to $\mathcal{O}\left(\alpha^5_s\left(\mathcal{T}^{\rm{cut}}\right)\right)$). This leads to the pattern of behaviour we see in figure \ref{fig:softDeviations}.

The small size of the coefficients for the NLL$^{\prime}$ variations suggests that the uncertainty is being underestimated here, whereas the $\mathcal{O}(1)$ coefficients for the NNLL$^{\prime}$ variations suggests that the uncertainty estimate here should be more trustworthy.

\subsection{\label{sec:channelcancellation}Parton Channel Cancellations}

In this section we study how the individual partonic channels contribute to the scale profile variations. We define a partonic channel by the initial partons from the protons as described by the PDFs. At NLL$^\prime$ + NLO the available channels are $q\bar{q}$ and $qg$. At NNLL$^\prime$ +NNLO there are more channels ($gg$, $qq'$ and $qq$) but the dominant channels are the same as NLL$^\prime$ + NLO and therefore are the focus of the following discussion. 

\begin{figure}[htbp]
\centering
\hspace{-2mm}
\includegraphics[width=.5\textwidth]{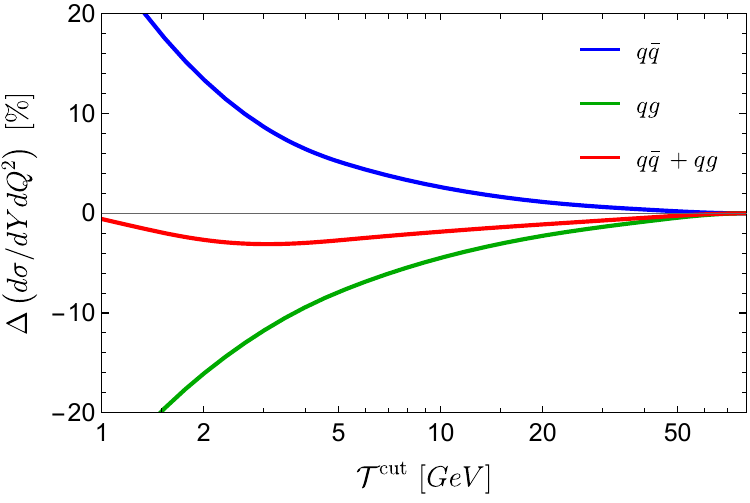}
\hspace{-2mm}
\includegraphics[width=.5\textwidth]{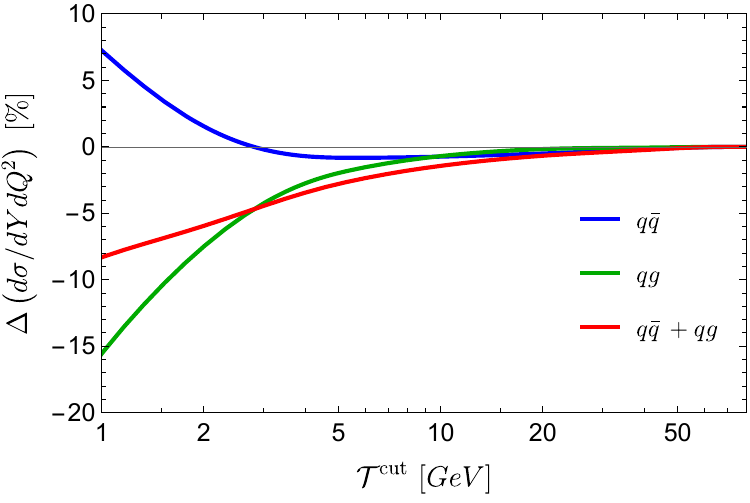}
\caption{Percent deviations in $q\bar{q}$ and $qg$ channels for NLL$^{\prime}$ (left) and NNLL$^{\prime}$ (right) differential $\Tau_{Bj}~<~\Tau^{\text{cut}}$ cross section for the $\beta = 1/6$ variation.}
\label{fig:qqqgcancel}
\end{figure}

We produced the deviations for the ($\alpha$,$\beta$) variations applied individually to the $q\bar{q}$ and the $qg$ partonic channels. Note that the $\mu_B$ and $\mu_S$ dependence is compensated between the beam/soft functions and the resummation factors within a partonic channel, so from the point of view of these scale variations this is a well-defined procedure. However we should not vary $\mu_F$ only for only one channel, since DGLAP evolution mixes the parton channels -- thus, even for these individual variations we vary $\mu_F$ in all channels together. The results for the case of the $\beta = 1/6$ deviation are given in figure \ref{fig:qqqgcancel}. These results are given for the differential cross section at $Y=0$ and $Q^2 = M_Z^2$; the same behaviour is seen in the integrated cross section. 

In figure \ref{fig:qqqgcancel} one observes that at NLL$^{\prime}$, the variations for the partonic channels have the opposite sign, and there is a large cancellation between channels when performing the variation for both channels together. However, for NNLL$^{\prime}$ the individual channel variations have the same sign outside the nonperturbative region and add together. A similar pattern is observed for the other ($\alpha$,$\beta$) variations -- for some variations, there is some cancellation between channels at NNLL$^{\prime}$, although never as much as for the NLL$^{\prime}$ case. The large cancellations in the NLL$^{\prime}$ case are accidental and will lead to an underestimation of the theoretical uncertainty when using the overall ($\alpha$,$\beta$) variations. A promising feature of the ($\alpha$,$\beta$) variations for the individual channels is that the NNLL$^{\prime}$ deviation is smaller than the NLL$^{\prime}$ deviation as expected. It was explicitly checked that this type of cancellation does not occur in the prediction for the gluon-fusion Higgs + 0-jet predictions with a  $\mathcal{T}_{B/Cj}$ veto from \cite{Gangal:2020qik}; here the $gg$ channel simply dominates the uncertainty prediction and there is no such cancellation.

Given this behaviour, we decided to produce results with a modified scale variation, that we shall refer to as \textit{MaxDev}. Here the resummation uncertainty is obtained by performing the individual channel variations for the $q\bar{q}$ and $qg$ channels, and then picking the largest deviation out of the two. We used this prescription for the final results in section \ref{sec:results}. Note that even with this prescription we anticipate that the resummation uncertainty for the NLL$^{\prime}$ results will be underestimated, due to the effects described in section \ref{sec:softErr}.

\subsection{\label{sec:Nuisance Parameters}Missing Higher Order Terms and Nuisance Parameters}

In this section, we aim to use an alternative method to produce another estimate of the theoretical uncertainties to check the \textit{MaxDev} approach gives reasonable values for the resummation uncertainties. To do this we will utilise the method of Theory Nuisance Parameters (TNPs), which is introduced and described in detail in ref. \cite{Tackmann:2024kci}. Section 6.2 of that paper discusses the implementation of the TNP approach in the context of a SCET resummation calculation with beam, soft and hard functions, which is what is needed here. Our aim here is to make a simple alternative estimate of the resummation uncertainty to check our existing \textit{MaxDev} estimate, and not to replace the latter, so we use a simplified version of the TNP method as outlined in the following sections.

Given that the issues identified earlier in this section are located in the resummation region, we focus on this region and do not include matching to the FO cross section. Further, to simplify and speed up the calculations, we will perform them for on-shell $Z$ boson production and only for the $\mathcal{T}_{Bj}$ observable; we expect the pattern of results to essentially be the same for Drell-Yan and for $\mathcal{T}_{Cj}$ (the main difference will be some overall normalisation difference of all uncertainty estimates due to the different Born level cross sections).

\subsubsection{\label{sec:TNP Overview}Overview of Method}

In our resummation formula (and many others), the ingredients are the fixed-order pieces ($H, B, S$) and the non-cusp anomalous dimensions $\gamma$ (plus the cusp anomalous dimension $\Gamma$). In order to increase the precision of the resummation from an  N$^{n}$LL$^\prime$ to N$^{n+1}$LL$^\prime$ we need these ingredients to the $n+1$-loop order (and the cusp anomalous dimension at $n+2$ loops). At a given order, much of the structure of each of the fixed-order pieces can be predicted from its renormalization group equation (RGE) and quantities already known for the N$^{n}$LL$^\prime$ calculation, such that a full $(n+1)$ loop calculation is only needed to determine the $\mu$-independent boundary conditions $F$, and the $(n+1)$-loop anomalous dimensions $\gamma$.

If we were applying the full TNP method of ref.~\cite{Tackmann:2024kci} to our N$^n$LL$^\prime$ prediction, we would first augment the calculation by adding all the pieces at the next order that can be determined from the RGE along with ingredients that were already known in the N$^n$LL$^\prime$ calculation, shifting the central prediction. We would then include nuisance parameters for the remaining boundary conditions $F$ and anomalous dimensions $\gamma$ and $\Gamma$, varying these by an appropriate amount to generate the theoretical uncertainty. In ref.~\cite{Tackmann:2024kci}, a proposal was made for an appropriate size of these variations by looking at the size of $F$, $\gamma$ and $\Gamma$ for various observables. Writing the expansions of $F$, $\gamma$ and $\Gamma$ as:
\begin{align}
F\left(\alpha_s\right) &= 1 + \sum_{n=1} \left(\frac{\alpha_s}{4\pi}\right)^n F_n,
\label{eq:FExp}
\\
\Gamma\left(\alpha_s\right) &= \sum_{n=0}\left(\frac{\alpha_s}{4\pi}\right)^{n+1}\Gamma_n,
\label{eq:GamExp}
\\
\gamma\left(\alpha_s\right) &= \sum_{n=0}\left(\frac{\alpha_s}{4\pi}\right)^{n+1}\gamma_n,
\label{eq:gamExp}
\end{align}
then ref.~\cite{Tackmann:2024kci} suggests that the nuisance parameter for $F_{n+1}$, $\gamma_{n+1}$ and $\Gamma_{n+1}$ should be:
\begin{align}
F_{n+1}\left(\theta_{n+1}\right) =&4C_{r}\left(4C_A\right)^{n}\, n!\,\,\theta_{n+1},
\label{eq:FNorm}
\\
\Gamma_{n+1}\left(\theta_{n+1}\right) =& 2C_{r}\left(4C_A\right)^{n+1}\theta_{n+1},
\label{eq:GamNorm}
\\
\gamma_{n+1}\left(\theta_{n+1}\right) =& 4C_{r}\left(4C_A\right)^{n+1}\theta_{n+1}.
\label{eq:gamNorm}
\end{align}
In this equation, $C_r$ is the leading colour factor (so, for example $C_r = C_F$ for the the Drell-Yan soft, hard, and $I_{qq}$ matching coefficient, whilst $C_r = T_F$ for the $I_{qg}$ matching coefficient), and $\theta_{n+1}$ is varied by an amount of $\mathcal{O}(1)$. For beam functions one varies it between $\pm 1$, for (colour-singlet production) hard functions one does the same but at the level of the Wilson coefficient $C$, and for the dijet soft functions one varies it between $\pm 2$ (roughly speaking, this is because the single soft function `talks to' two beam functions). The total uncertainty is determined by adding the uncertainties from individual TNPs in quadrature.

Here we do not want to shift our central N$^n$LL$^\prime$ predictions, but rather to generate an alternative uncertainty band for these predictions. Thus, we only include and vary the TNP terms for the boundary terms and anomalous dimensions, omitting the step where the terms determined from the RGE are included. In essence, we are then simply taking the fundamental pieces of the resummation formula at N$^n$LL$^\prime$, and adding a term which is a coefficient multiplied by $\alpha_s^{n+1}$ to each (or $\alpha_s^{n+2}$ in the case of $\Gamma$), which is supposed to represent the collection of terms appearing at the next order in perturbation theory. Each coefficient is then varied by an appropriate amount centred on zero to determine the uncertainty band. We did check at the NLL$^\prime$ level that the `shift' in the central prediction from including the terms determined from the RGE was in any case smaller than the final TNP uncertainty band presented below, and expect the same to hold true at NNLL$^\prime$. Note that since we do not want to shift the central predictions, we also maintain the N$^n$LO PDFs and central N$^n$LL$^\prime$ profile scales in our N$^n$LL$^\prime$ predictions.

\subsubsection{\label{sec:TNP NLL}NLL$^\prime$ Implementation}

The TNPs we need here are boundary constants related to the 2-loop factorization functions $\{I^{\left(c\right)}_{qq},I^{\left(c\right)}_{qg},S^{\left(c\right)},H^{\left(c\right)}\}$ and the anomalous dimensions $\{\Gamma^{q}_2,\gamma^{q}_{S1},\gamma^{q}_{B1}\}$ (note that although we write $H^{\left(c\right)}$, the TNP is implemented at the level of the Wilson coefficient). The $c$ superscript indicates that these are boundary conditions. We do not have a nuisance parameter for the hard anomalous dimension $\gamma_H$ as it is set by RG consistency. Several of these quantities are actually functions of the jet radius $R$, but here we simply fix $R=0.5$ and do not concern ourselves with the $R$-dependence. The beam function matching coefficients $I^{\left(c\right)}_{qq}$ and $I^{\left(c\right)}_{qg}$ should also be functions of $z$, but we also ignore this for the corresponding TNPs, taking them to be constant in $z$. 

At NLL$^\prime$ we actually know how big the TNP variations `should be' since we have the actual results for the NNLL$^\prime$ resummation ingredients. We can adjust the range of variation of each TNP from the `default' of section \ref{sec:TNP Overview} such that its maximum magnitude corresponds to the actual value of the NNLL$^\prime$ ingredient. For the $I^{\left(c\right)}_{qq/qg}$ matching coefficients, we adjust the maximum size of the TNP so that the convolution of this with the PDFs is of the same size as the convolution of the actual NNLO matching coefficient with the PDF, for typical $x$ values probed in the Drell-Yan process. At this point we are of course forcing the TNP variations to be reasonable and cover the difference between the NLL$^\prime$ and NNLL$^\prime$ predictions `by hand' using information from the NNLL$^\prime$ order. The idea here is to take some lessons from this exercise to use at NNLL$^\prime$ (where we don't know the ingredients at the next order), and it is also interesting to compare these variations to the `default' ones discussed below equations \eqref{eq:FNorm}, \eqref{eq:GamNorm} and \eqref{eq:gamNorm}.

This comparison is shown in table \ref{tab:scaling}, where the TNP ranges set by the true NNLL$^\prime$ coefficients are denoted by `true value variation'. One noteworthy aspect is that the `true' variation of the $I^{(c)}_{ij}$ matching coefficient TNPs is different from $\pm 1$; this is perhaps expected as the approach of taking these to be independent of $z$ is somewhat of an oversimplification. However, we also observe that the TNP variations for the soft boundary constant and the non-cusp anomalous dimensions, $\gamma_{S1}^q$, $\gamma_{B1}^q$ and $S^{\left(c\right)}$, are $\simeq 3$ times larger than the default prescriptions. For the anomalous dimensions, this is caused by the fact that the jet radius clustering corrections at $\mathcal{O}(\alpha_s^2)$ are quite large (as was previously noted in the context of $p_{Tj}$ in ref.~\cite{Alioli:2013hba}). For the soft boundary constant and anomalous dimensions, it was found that for the ``global'' piece and the clustering corrections individually (i.e. $S_G/\gamma_G$ and $\Delta S/\Delta \gamma$ in ref. \@\cite{Gangal:2016kuo}), their size corresponded fairly well to the default variations of equations \eqref{eq:FNorm} and \eqref{eq:gamNorm}. Thus, if one were to take an approach with separate TNPs for the global and clustering pieces, one would not see such a strong enhancement of the `true' variation over the default one.

\begin{table}
\centering
\begin{tabular}{|c|c|c|c|}
    \hline
    TNP & True value variation & Default variation &  Ratio of variations\\
    \hline
    $\Gamma^{q}_2$ & $\pm0.5$ & $\pm 1.0$& $0.50$\\
    \hline
    $\gamma^{q}_{S1}$ & $\pm5.5$&$\pm2.0$ & $2.75$\\
    \hline
    $\gamma^{q}_{B1}$ & $\pm3.0$ & $\pm1.0$&$3.00$\\
    \hline
    $I^{\left(c\right)}_{qq}$ & $\pm6.0$ &$\pm1.0$ &$6.00$\\
    \hline
    $I^{\left(c\right)}_{qg}$ & $\pm0.3$ &$\pm1.0$ &$0.30$\\
    \hline
    $S^{\left(c\right)}$ & $\pm6.0$ &$\pm2.0$ &$3.00$\\
    \hline
    $H^{\left(c\right)}$ & $\pm1.0$ &$\pm1.0$ &$1.00$\\
    \hline
\end{tabular}
\caption{\label{tab:scaling} Different approaches for setting the TNP ranges at NLL$^\prime$. The column labelled `true value variation' shows the  $\theta_2$ values when these are set using the NNLL$^\prime$ ingredients, whilst the `default' variations are the ones of ref.\@ \cite{Tackmann:2024kci}. In the final column we give the ratio of the `true' variations to the default ones.}
\end{table}

\begin{figure}
\centering
\hspace{-2mm}
\includegraphics[width=.5\textwidth]{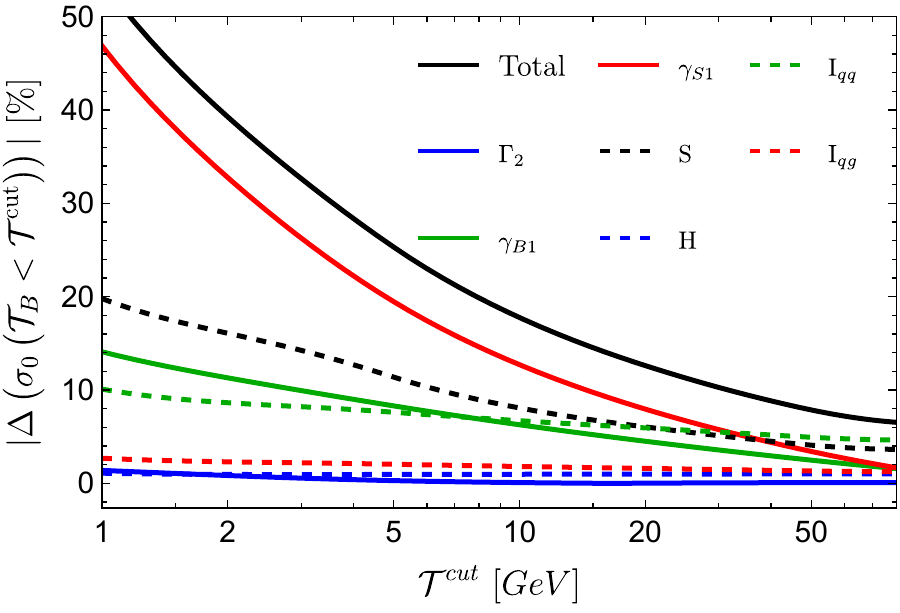}
\hspace{-2mm}
\includegraphics[width=.5\textwidth]{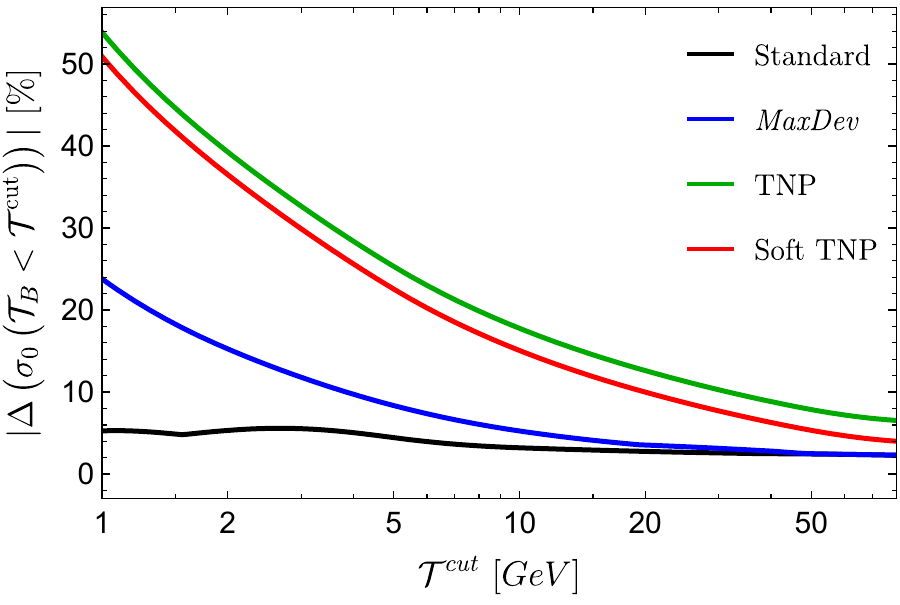}
\caption{Predictions for the NLL$^\prime$ $\mathcal{T}_{Bj} \leq \mathcal{T}^{\text{cut}}$ theory uncertainty using theory nuisance parameters for the on-shell $Z$ boson production cross section. The predictions are shown split by individual nuisance parameter contribution (left) and compared with scale variation uncertainties (right).}
\label{fig:NLLTNP}
\end{figure}

Figure \ref{fig:NLLTNP} shows the breakdown of the NLL$^\prime$ TNP uncertainty prediction (using the `true value variations' from table \ref{tab:scaling}) as well as the comparison with the NLL$^\prime$ scale variation theory uncertainties. We first note that the TNPs for the soft ingredients $S^{(c)}$ and $\gamma_{S1}$ dominate the TNP uncertainty in the resummation region; in the right hand panel of figure \ref{fig:NLLTNP} we include a curve that only takes account of these TNPs (`soft TNP'), and we see that it captures the bulk of the total TNP uncertainty. 
The \textit{MaxDev} uncertainty lies closer to the TNP uncertainty than the standard one does, and the \textit{MaxDev} and TNP uncertainties have a similar shape, but the TNP uncertainty is still substantially larger than either of the scale uncertainties. This is expected; we know that the \textit{MaxDev} procedure will still underestimate the uncertainty at NLL$^\prime$ due to the anomalously small effect of soft scale variations discussed in section \ref{sec:softErr}.

\subsubsection{\label{sec:TNP NNLL}NNLL$^\prime$ Implementation}

To construct the NNLL$^\prime$ TNP uncertainty estimate, we shall assume the size of the N$^3$LL$^\prime$ ingredients follows a similar pattern as observed for the NNLL$^\prime$ ones, so that we can use the variations in table \ref{tab:scaling}, and also just use the variations in the TNPs for the soft ingredients to get a reasonable estimate of the total TNP uncertainty. This is possibly a conservative estimate at NNLL$^\prime$ -- in ref.~\cite{Alioli:2013hba} (see also ref.~\cite{Dasgupta:2014yra}) it was observed that for the (rapidity) anomalous dimension for $p_{Tj}$, the leading $\log(R)$ jet clustering term is somewhat smaller at $\mathcal{O}(\alpha_s^3)$ than the corresponding term at $\mathcal{O}(\alpha_s^2)$, so if this pattern extends to $\Tau_{B/Cj}$ and the full jet clustering correction, we overestimate the uncertainty associated with $\gamma_{S2}$.

\begin{figure}
\centering
\includegraphics[width=.75\textwidth]{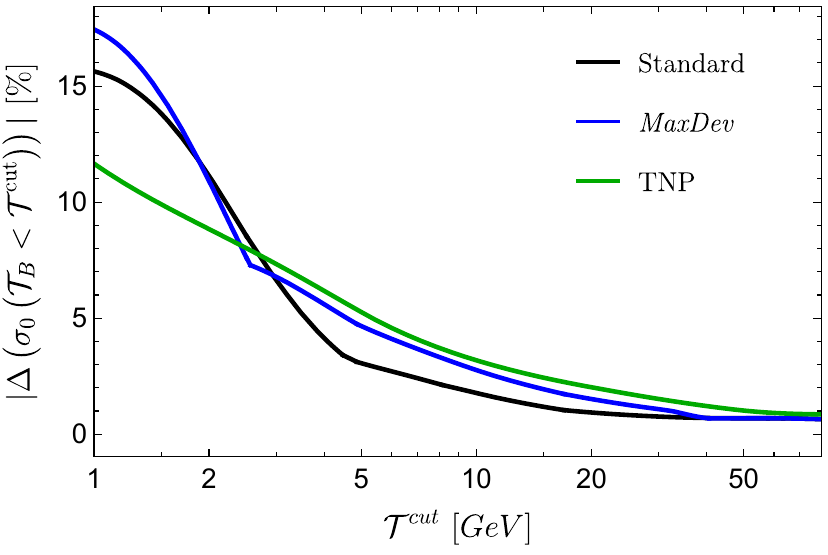}
\caption{Predictions for the NNLL$^\prime$ $\mathcal{T}_{Bj} \leq \mathcal{T}^{\text{cut}}$ theory uncertainty for the on-shell $Z$ boson production cross section, where the TNP uncertainty prediction uses only the soft TNPs and conservative variations.}
\label{fig:NNLLTNP}
\end{figure}

Figure \ref{fig:NNLLTNP} shows comparisons of the NNLL$^\prime$ theory uncertainty using scale variations and our (likely conservative) TNP approach. We observe that above the nonperturbative region, the \textit{MaxDev} and TNP uncertainty estimates agree fairly well, and are both larger than the standard scale variation uncertainty. This gives us some confidence that the \textit{MaxDev} uncertainty is reasonable at the NNLL$^\prime$ level.

\section{\label{sec:results}Resummed Predictions up to NNLL$^{\prime}$ + NNLO}

In section \ref{sec:mainResults} we present our final results for the Drell-Yan process $pp \to Z/\gamma^* + X \to l^+l^- + X$ in the presence of the $\Tau_{B/Cj}$ jet vetoes, at both NLL$^\prime$+NLO and NNLL$^\prime$+NNLO. These are compared to the fixed order NLO and NNLO results to assess in what region of $\Tau^\cut$ the resummation has an impact. In section \ref{sec:piEffects} we study the impact of the $\pi^2$ resummation (discussed in section \ref{sec:resummation}) on both our NLL$^\prime$+NLO and NNLL$^\prime$+NNLO results.

We first summarise the setup used in our analysis: we use the MSHT20nloas120 PDFs for the NLO and NLL$^{\prime}$ + NLO predictions, and the MSHT20nnloas118 PDFs for the NNLO and NNLL$^{\prime}$ + NNLO predictions \cite{Bailey:2020ooq}. These predictions were implemented in Mathematica \cite{Mathematica} using a PDF interpolation package created and described in ref.~\cite{Martin:2009iq}. The mass and width of the $Z$ boson were set to be $M_Z = 91.1876\text{ GeV}$ and $\Gamma_Z = 2.4952\text{ GeV}$ respectively. We give predictions for the production of massless lepton pairs of a single species (e.g. electron-positron pairs).  We consider the centre-of-mass energy to be 13 TeV, integrate $Q$ between  $80 - 100\, \text{  GeV}$ and use a jet radius $R$ of $0.5$. We take $n_f = 5$. In producing the results for $\Tau_{Bj}$ we take $r_s=1$ in eq.~\eqref{eq:frun}, whilst for $\Tau_{Cj}$ we take $r_s=2$, following ref.~\cite{Gangal:2020qik}.

\subsection{\label{sec:mainResults}Drell-Yan Jet Veto Predictions}

\begin{figure}
\centering
\hspace{-2mm}
\includegraphics[width=.5\textwidth]{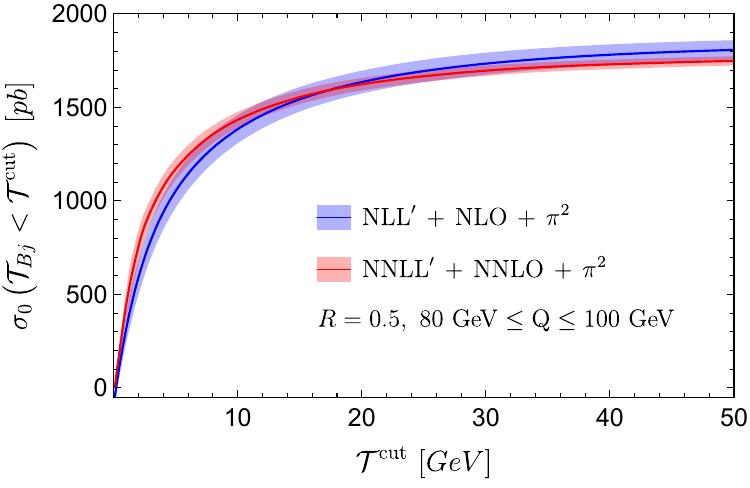}
\hspace{-2mm}
\includegraphics[width=.5\textwidth]{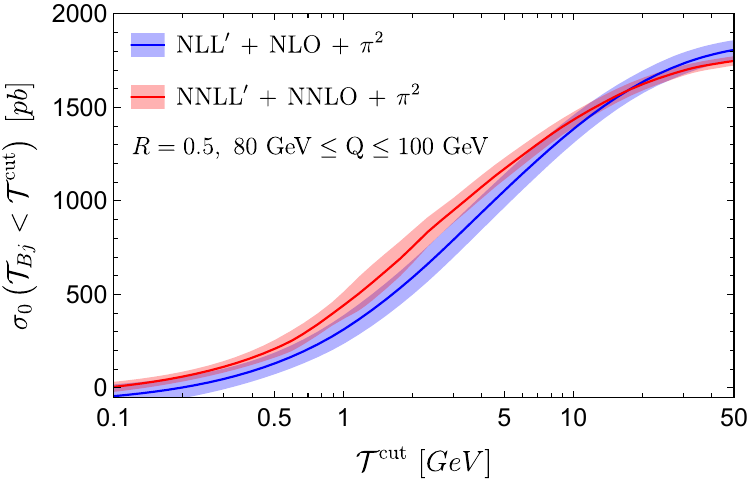}
\caption{Drell-Yan NLL$^\prime$ + NLO (blue) and NNLL$^\prime$ + NNLO (red) cross section for $\Tau_{Bj} < \Tau^\mathrm{cut}$ plotted on a linear scale (left) and logarithmic scale (right) using the \textit{MaxDev} prescription for the uncertainty.}
\label{fig:tauBmax}
\end{figure}

\begin{figure}
\centering
\hspace{-2mm}
\includegraphics[width=.5\textwidth]{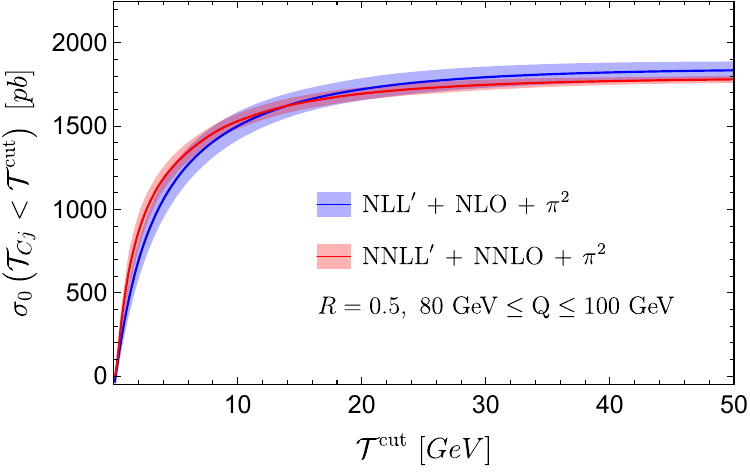}
\hspace{-2mm}
\includegraphics[width=.5\textwidth]{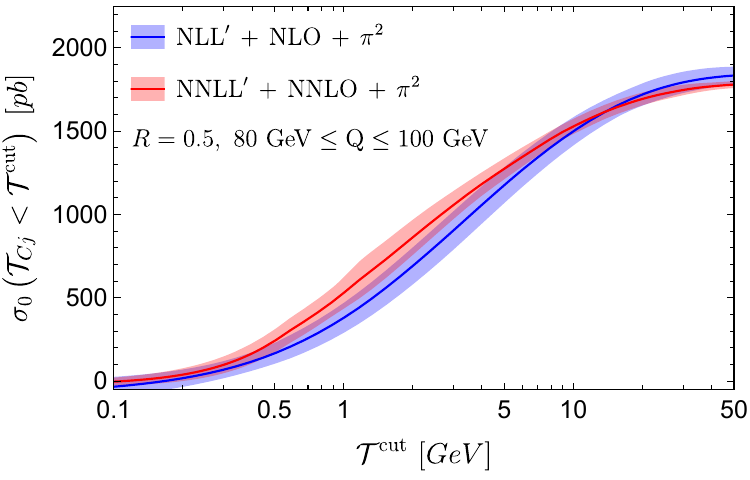}
\caption{Drell-Yan NLL$^\prime$ + NLO (blue) and NNLL$^\prime$ + NNLO (red) cross section for $\Tau_{Cj}~<~\Tau^\mathrm{cut}$ with $r_s = 2 $ plotted on a linear scale (left) and logarithmic scale (right) using the \textit{MaxDev} prescription for the uncertainty.}
\label{fig:tauCmax}
\end{figure}

We present in figure \ref{fig:tauBmax} and figure \ref{fig:tauCmax} our predictions for the NLL$^{\prime}$ + NLO and NNLL$^{\prime}$ + NNLO Drell-Yan cross section, for $\Tau_{Bj} < \Tau^\mathrm{cut}$ and $\Tau_{Cj} < \Tau^\mathrm{cut}$ respectively. We use the \textit{MaxDev} prescription for the resummation uncertainty as discussed in section \ref{sec:channelcancellation}. One observes that, under this prescription, the NNLL$^{\prime}$ uncertainty band is smaller than the NLL$^{\prime}$ band (above the `nonperturbative region' as defined in section \ref{sec:uncertainty}), and the bands overlap (unlike the `standard' uncertainty bands in figure \ref{fig:tauBstnd}). Note that, as discussed in section \ref{sec:convergence}, the theoretical uncertainties at NLL$^{\prime}$ + NLO are likely still under predicted, whilst those at NNLL$^{\prime}$ + NNLO should be reasonable.

\begin{figure}[htbp]
\centering
\hspace{-2mm}
\includegraphics[width=.47\textwidth]{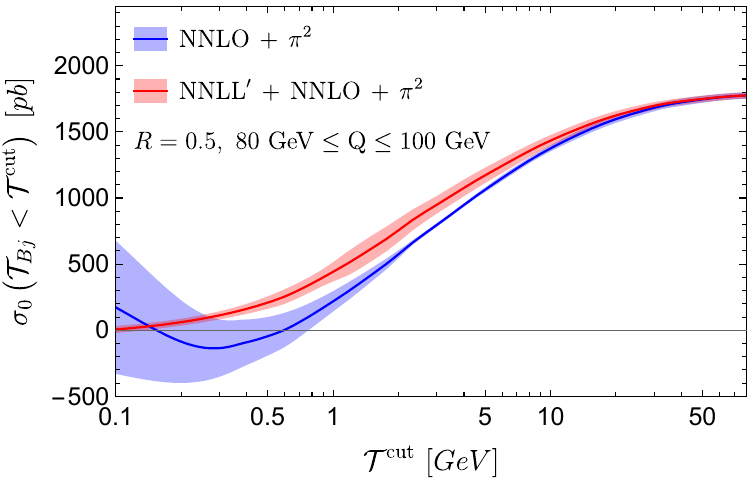}
\hspace{-2mm}
\includegraphics[width=.47\textwidth]{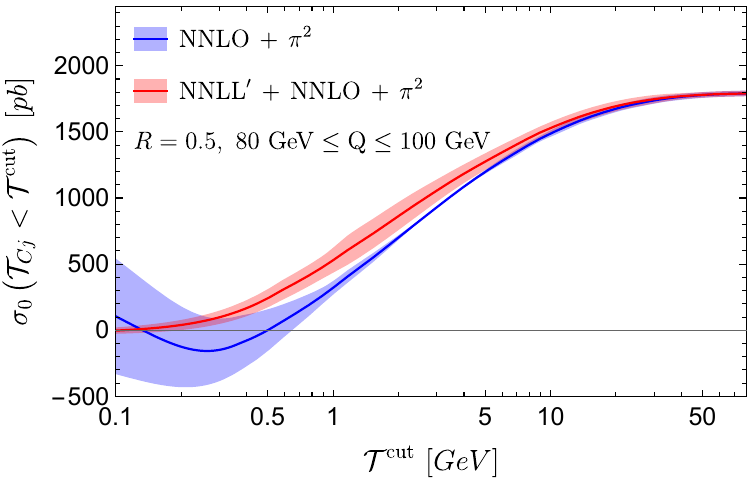}
\caption{Drell-Yan NNLL$^{\prime}$ + NNLO resummed predictions compared with fixed-order NNLO predictions for $\Tau_{Bj} < \Tau^{\text{cut}}$ (left) and $\Tau_{Cj} < \Tau^{\text{cut}}$ (right).}
\label{fig:tauFOvMax}
\end{figure}

We compare our resummed predictions with fixed-order NNLO $\pi^2$-improved cross section predictions in figure \ref{fig:tauFOvMax} for $\Tau_{Bj}$ and $\Tau_{Cj}$ vetoes. The uncertainty bands on the NNLO predictions are 
simply obtained by the variation of $\mu_{\rm{FO}}$ by a factor of $2$ around $M_Z$, so we do not expect them to be reliable for $\Tau^\cut \ll M_Z$. We see that the central predictions for NNLO and fixed NNLL$^\prime$+NNLO differ appreciably below around $20$ GeV -- this is roughly consistent with figures \ref{fig:TauBScaleProfile} and \ref{fig:TauCScaleProfile}, where the singular contributions dominate in this region (recall that we set the end of the resummation region to be $\sim 18$ GeV). At very small values of $\Tau^\cut$, the fixed order prediction actually becomes completely unphysical ($<0$), whilst our resummed prediction goes to zero as expected. At large $\Tau^\cut$, we find that our matched NNLL$^{\prime}$ + NNLO predictions precisely coincide with the fixed order ones, as required.

In table~\ref{tab:results}, we present cross section values in the presence of a $\Tau_{B/Cj}$ veto, for $\Tau^\cut = 10,20,30$ GeV, at various fixed order and resummed accuracies. For the resummed predictions, both the standard and \textit{MaxDev} uncertainties are given. The theoretical uncertainty in our NNLL$^{\prime}$+NNLO results is in the range of $1-3\%$, which is comparable to that for state-of-the-art predictions for Drell-Yan production with a veto on the $p_T$ of jets \cite{Banfi:2012yh,Banfi:2012jm,Becher:2014aya,Campbell:2023cha}. The experimental uncertainty in a recent measurement of Drell-Yan $Z$ production with a veto $p_{Tj} < 30$ GeV imposed is $\sim 2.75\%$ \cite{CMS:2022ilp}. Although the measurement
for Drell-Yan with a veto using $\Tau_{B/Cj}$ observables is currently not available, one would anticipate a similar level of uncertainty in the experimental measurement, and hence our predictions should be sufficiently precise for detailed comparisons. 

\begin{table}
\centering
\begin{tabular}{|c|c|c|}
    \hline
    & $\sigma_0\left(\Tau_{B_j} < \Tau^{\text{cut}}\right)\left[\text{pb}\right]\left(r_s=1\right)$ & $\sigma_0\left(\Tau_{C_j} < \Tau^{\text{cut}}\right)\left[\text{pb}\right]\left(r_s=2\right)$ \\
    \hline
    NLL$^{\prime}$ + NLO & & \\
    $\Tau^{\text{cut}} = 10\text{ GeV}$ & $1384 \pm 46 \left(3.3\%\right)$ & $1500 \pm 51 \left(3.4\%\right)$\\
    $\Tau^{\text{cut}} = 20\text{ GeV}$ & $1636 \pm 47 \left(2.9\%\right)$ & $ 1721 \pm 53  \left(3.1\%\right)$ \\
    $\Tau^{\text{cut}} = 30\text{ GeV}$ & $1734 \pm 48 \left(2.8\%\right)$ & $1793 \pm 52 \left(2.9\%\right)$\\
    \hline
    NNLL$^{\prime}$ + NNLO & & \\
    $\Tau^{\text{cut}} = 10\text{ GeV}$ & $1434 \pm 26 \left(1.8\%\right)$ & $1529 \pm 35 \left(2.3\%\right)$\\
    $\Tau^{\text{cut}} = 20\text{ GeV}$ & $1624 \pm 26 \left(1.6\%\right)$ & $ 1694 \pm 28  \left(1.7\%\right)$ \\
    $\Tau^{\text{cut}} = 30\text{ GeV}$ & $1697 \pm 24 \left(1.4\%\right)$ & $1747 \pm 25 \left(1.4\%\right)$\\
    \hline
    NLL$^{\prime}$ + NLO \textit{MaxDev} & & \\
    $\Tau^{\text{cut}} = 10\text{ GeV}$ & $1384 \pm 75 \left(5.4\%\right)$ & $1500 \pm 85\left(5.7\%\right)$ \\
    $\Tau^{\text{cut}} = 20\text{ GeV}$ & $1636 \pm 60 \left(3.7\%\right)$ & $ 1721 \pm 67  \left(3.9\%\right)$ \\
    $\Tau^{\text{cut}} = 30\text{ GeV}$ & $1734 \pm 59 \left(3.4\%\right)$ & $1793 \pm 64 \left(3.6\%\right)$ \\
    \hline
    NNLL$^{\prime}$ + NNLO \textit{MaxDev} & & \\
    $\Tau^{\text{cut}} = 10\text{ GeV}$ & $1434 \pm 42 \left(2.9\%\right)$ & $1529 \pm 46 \left(3.0\%\right)$ \\
    $\Tau^{\text{cut}} = 20\text{ GeV}$ & $1624 \pm 33 \left(2.0\%\right)$ & $ 1694 \pm 38  \left(2.2\%\right)$ \\
    $\Tau^{\text{cut}} = 30\text{ GeV}$ & $1697 \pm 28 \left(1.6\%\right)$ & $1747 \pm 31 \left(1.8\%\right)$ \\
    \hline
\end{tabular}
\caption{\label{tab:results}Numerical results for various resummed predictions of the $\Tau_{B/Cj} < \Tau^{\text{cut}}$ Drell-Yan cross section.}
\end{table}

\subsection{\label{sec:piEffects}Effects of $\pi^2$ Resummation}
In this section we explore the effects of the inclusion of $\pi^2$ resummation discussed in section \ref{sec:uncertainty}. To investigate this, we obtained predictions for the $\Tau_{Bj} < \Tau^{\rm cut}$ cross section with and without the inclusion of $\pi^2$ resummation. For simplicity, we perform these calculations for on-shell $Z$ boson production, expecting the pattern of results to be similar for Drell-Yan (as we did in section \ref{sec:Nuisance Parameters}). The comparison should also be very similar for $\Tau_{Cj}$ so we do not present this here. Predictions were produced with theory uncertainties obtained using both the standard and \textit{MaxDev} prescriptions. The effect of adding the $\pi^2$ resummation was the same for both uncertainty prescriptions and therefore only the \textit{MaxDev} results will be discussed below.

\begin{figure}[htbp]
\centering
\hspace{-2mm}
\includegraphics[width=.5\textwidth]{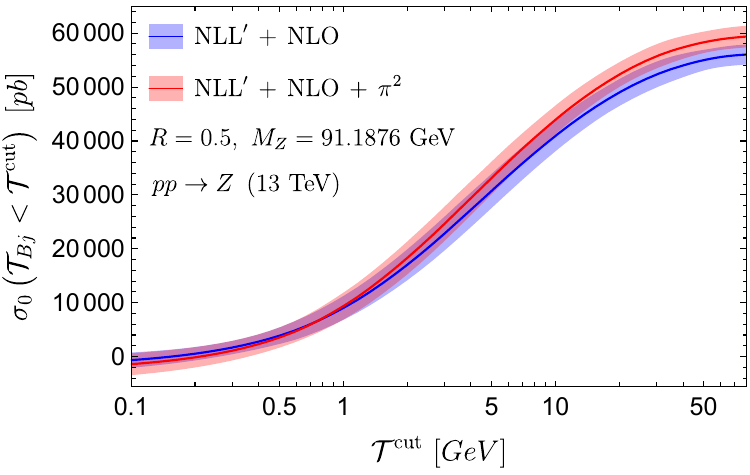}
\hspace{-2mm}
\includegraphics[width=.5\textwidth]{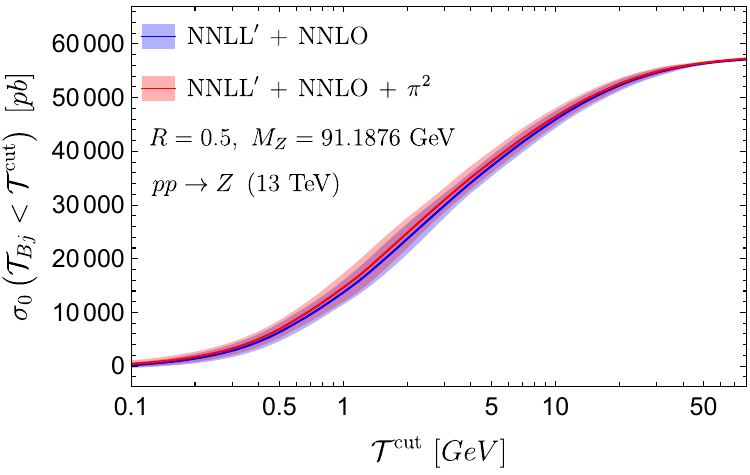}
\caption{Resummed predictions with and without the inclusion of $\pi^2$ resummation for the on-shell $Z$ production cross section with a veto $\Tau_{Bj} < \Tau^{\rm cut}$. The NLL$^\prime$ + NLO predictions (left) and NNLL$^\prime$ + NNLO predictions (right) are plotted on a logarithmic scale.}
\label{fig:pivNopi}
\end{figure}

Figure \ref{fig:pivNopi} shows the NLL$^\prime$ + NLO and NNLL$^\prime$ + NNLO predictions for on-shell $Z$ boson production with and without the inclusion of $\pi^2$ resummation. Looking first at the NLL$^\prime$ + NLO prediction, the size of the uncertainties is relatively unchanged (as might be expected), but there is a clear upwards shift of the central prediction, especially towards larger $\Tau^\cut$ values. By contrast, the NNLL$^\prime$ + NNLO predictions do not show any noticeable change due to $\pi^2$ resummation with the uncertainty bands from both predictions being on top of each other. Thus, for our highest-precision NNLL$^\prime$ + NNLO predictions, the $\pi^2$ resummation does not have a noticeable impact.

\subsection{\label{sec:fiducial}Predictions with Fiducial Cuts}
\begin{figure}
\centering
\hspace{-2mm}
\includegraphics[width=.5\textwidth]{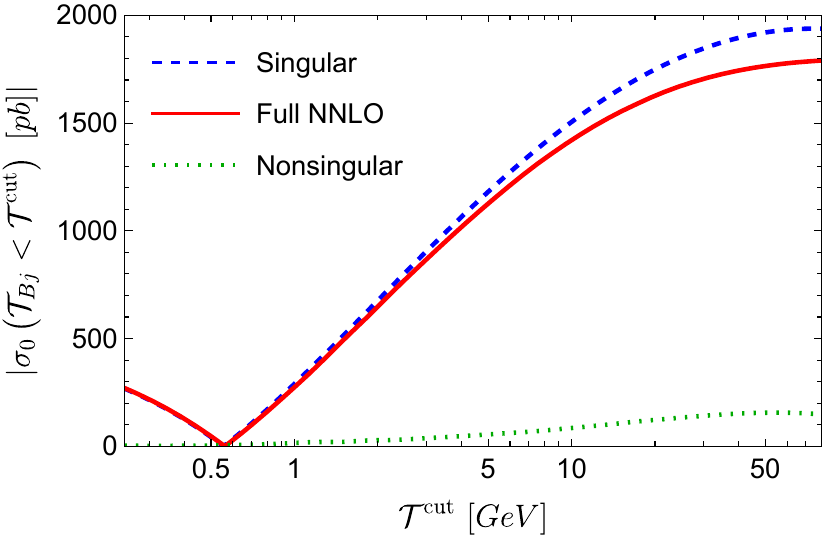}
\hspace{-2mm}
\includegraphics[width=.5\textwidth]{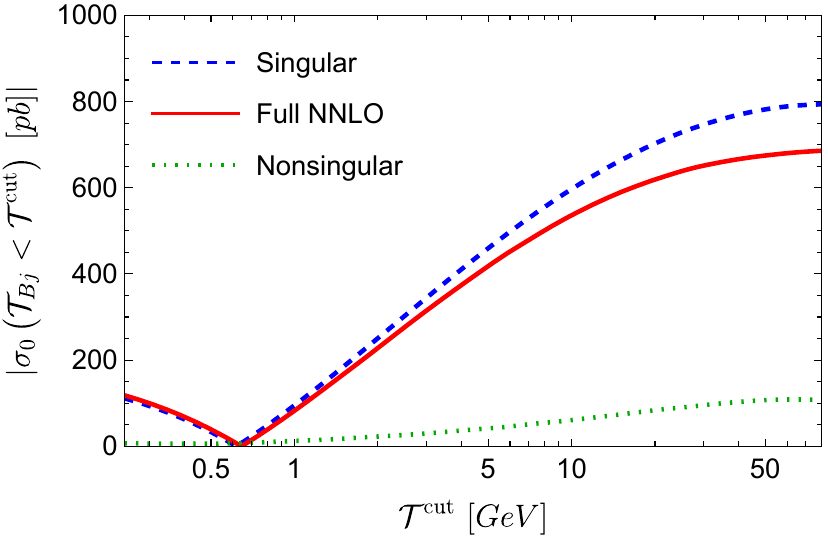}
\caption{The NNLO $\mathcal{T}_{Bj}<\mathcal{T}^{\rm{cut}}$ Drell-Yan cross section split into its singular and non-singular components. On the left we give results inclusive in the kinematics of the leptons, and the results on the right are in the presence of the cuts $p^l_T \ge 25 \text{ GeV}$ and $|\eta_l| \le 2.4$.
}
\label{fig:IvFNNLO}
\end{figure}

\begin{figure}
\centering
\hspace{-2mm}
\includegraphics[width=.5\textwidth]{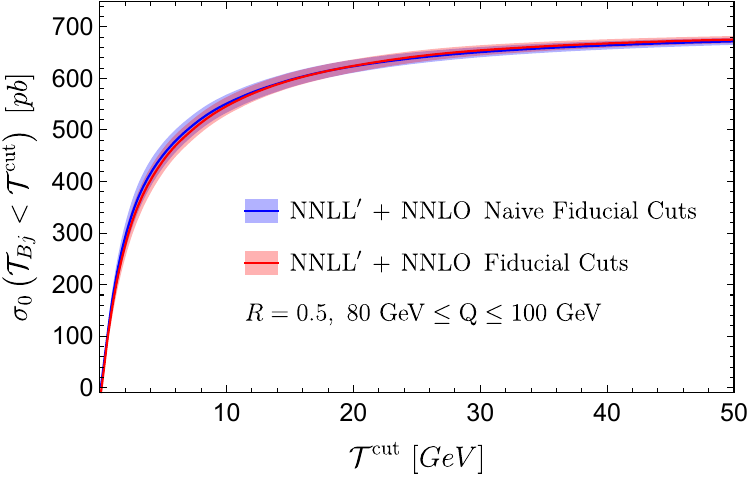}
\hspace{-2mm}
\includegraphics[width=.5\textwidth]{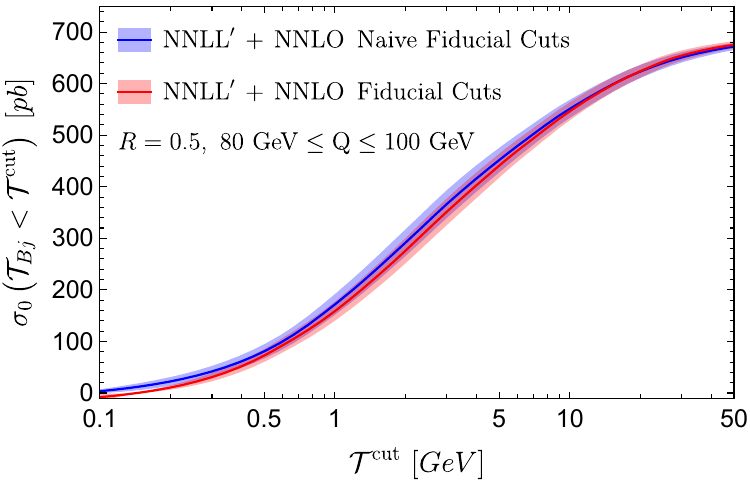}
\caption{Drell-Yan NNLL$^\prime$ + NNLO cross section for $\Tau_{Bj} < \Tau^\mathrm{cut}$ in the presence of the cuts $p^l_T \ge 25 \text{ GeV}$ and $|\eta_l| \le 2.4$. Plotted on a linear scale (left) and logarithmic scale (right).}
\label{fig:fiducial}
\end{figure}

The predictions presented so far are inclusive over the kinematics of the final state leptons. In practice, fiducial cuts will commonly be present in experimental
predictions to allow for a clean signature. We can account for these cuts by altering the Born cross section in the hard function of the resummed piece (first term in \eqref{eq:fullXsec}) and by adding such cuts to all terms in the nonsingular piece calculated using MADGRAPH5 AMC@NLO and DYTurbo. To demonstrate this, we produced the NNLL$^\prime$ + NNLO $ \sigma_{0}\left(\mathcal{T}_{Bj} \leq \mathcal{T}^{\rm{cut}}\right)$ with the following fiducial cuts on the leptons: $p^l_T~\ge~25~\text{ GeV}$ and $|\eta_l|~\le~2.4$. These are the same cuts used in the CMS Drell-Yan measurement in ref.~\cite{CMS:2019raw}. Here we do not include $\pi^2$ resummation due to the minimal impact at NNLL$^\prime$ + NNLO as seen in section \ref{sec:piEffects}. We first check the behaviour of the non-singular corrections is as expected. Figure \ref{fig:IvFNNLO} shows the absolute size of the singular and non-singular contributions to the NNLO cross section inclusive to lepton cuts and in the presence of fiducial cuts. It can clearly be seen the overall behaviour of the non-singular prediction is very similar between our inclusive and fiducial calculations, going to $0$ as $\mathcal{T}^{\rm{cut}}$ goes to $0$ as expected.
Figure \ref{fig:fiducial} shows the NNLL$^\prime$ + NNLO predictions for $ \sigma_{0}\left(\mathcal{T}_{Bj} \leq \mathcal{T}^{\rm{cut}}\right)$, where we compare the implementation just described with a very naive approach in which the NNLL$^\prime$+NNLO prediction inclusive in the lepton kinematics is simply rescaled by the ratio of the NNLO fiducial and inclusive cross sections (with the cross sections in this ratio having no $\Tau_{Bj}$ cut). It can be seen that the naive implementation is consistent with the full implementation in all regions of interest. The uncertainty bands in this plot are simply obtained by using the percent uncertainties from the soft TNP approach as seen in figure \ref{fig:NNLLTNP} -- this should be reasonable at small $\Tau^\cut$. These predictions are presented to demonstrate the fiducial implementation; predictions for $\mathcal{T}_{B/Cj}$ with the full uncertainty implementation  or different fiducial cuts can be provided on request.

\section{\label{sec:conclusion}Conclusions}

The Drell-Yan process, $pp \rightarrow Z/\gamma^* + X \rightarrow l^+ l^- + X$, is a crucial standard candle process at the LHC, and provides a clean testing ground for our understanding of perturbative QCD. As such, precise predictions for this process with a variety of different measurements applied to the accompanying QCD radiation are highly desirable. Here, we have obtained both NLL$^\prime$ + NLO and NNLL$^\prime$ + NNLO predictions for the Drell-Yan cross section with a cut on the rapidity-dependent jet veto observables $\Tau_{Bj}$ and $\Tau_{Cj}$. The $\Tau_{B/Cj}$ observables impose a veto on the transverse momentum of the leading jet weighted by the rapidity of the jet, such that the veto is tight at central rapidities and becomes looser at forward rapidities. Such vetoes have a somewhat different resummation structure than the `standard' $p_{Tj}$ veto, and they are convenient for use at the LHC, where missing tracking information at forward rapidities makes it difficult to impose a tight veto in this region. 

Initially we determined the uncertainty in the resummed predictions following the procedure described in ref.~\cite{Gangal:2020qik}, where that paper obtains similar predictions but in the context of Higgs production. We observed unsatisfactory behaviour in the uncertainty bands, with the NLL$^\prime$ + NLO band being smaller than the NNLL$^\prime$ + NNLO band at very small $\Tau^\cut$ values, and the bands not overlapping in this region. To establish the reasons for this, we examined the resummation scale variations, where the relevant scales are the soft scale $\mu_S$ and beam scale $\mu_B$. We found that the uncertainty for the NLL$^\prime$ + NLO prediction in particular was being underestimated. The reasons behind this were two-fold: first, there is an anomalously small response of the resummed cross section to variations of the soft scale $\mu_S$ at NLL$^\prime$, and second, there is a strong cancellation of the effect of resummation scale variations between partonic channels at this order. This cancellation also occurred at NNLL$^\prime$ although to a rather lesser degree. To alleviate the cancellation of uncertainty between partonic channels, we developed an alternative `\textit{MaxDev}' approach for the uncertainties where we only vary the beam scale $\mu_B$ in the partonic channel that yields the largest uncertainty band. We compared the results of this against a simplified version of the Theory Nuisance Parameter approach \cite{Tackmann:2024kci} and found agreement between the two for the NNLL$^\prime$ + NNLO case, suggesting that the uncertainty estimate is reasonable here. In the NLL$^\prime$ + NLO case the \textit{MaxDev} uncertainty is somewhat smaller than the TNP one, which is due to the fact that the former still suffers from the anomalously small response to $\mu_S$, and thus is still likely an underestimate.

We produced predictions with $R=0.5$ and for $Q$ integrated between 80 and 100 GeV (although results for other values of $R$ and/or $Q$ can be obtained on request from the authors). The resummation of time-like logarithms (`$\pi^2$ resummation') was included but was found to have a minimal effect on the NNLL$^{\prime}$ + NNLO predictions. Using the \textit{MaxDev} prescription for the uncertainties, we found that for the $\Tau_{Bj}$ observable and $\Tau^\cut = 10$ GeV, the uncertainty reduces from 5.6\% at NLL$^{\prime}$ + NLO to 3.1\% at NNLL$^{\prime}$ + NNLO, with the central value increasing from 1372 pb to 1439 pb. For the $\Tau_{Cj}$ case, the corresponding uncertainties are 5.8\% and 3.4\%, with the central values being 1486 pb and 1538 pb respectively. Predictions including fiducial cuts on the final state leptons were also produced, and compared to a naive approach in which the  NNLL$^\prime$+NNLO prediction inclusive in the lepton kinematics was simply rescaled by the ratio of the NNLO fiducial to NNLO inclusive cross sections; these predictions were found to be consistent (further results involving fiducial cuts can be obtained on request from the authors).

Our predictions for Drell-Yan production with a $\Tau_{B/Cj}$ veto are of a comparable precision to the corresponding state-of-the-art predictions with a $p_{Tj}$ veto, and it will be interesting to compare these with the experimental data from the LHC in the future.

\acknowledgments
JRG wishes to thank Thomas Cridge and Frank Tackmann for useful discussions on the topic of Theory Nuisance Parameters. SG acknowledges support from the Department of Science and Technology (DST), Government of India, under Grant No. IFA22-PH 296 (INSPIRE Faculty Award). The work of JRG, and part of the work of TC, has been supported by the Royal Society through Grant URF\textbackslash{}R1\textbackslash{}201500. TC also acknowledges financial support from a University of Manchester PGRTA scholarship.

\providecommand{\href}[2]{#2}\begingroup\raggedright\endgroup

\end{document}